\documentclass[prd, aps, nofootinbib, preprintnumbers, showpacs, superscriptaddress,twocolumn]{revtex4}

\usepackage{graphicx}
\usepackage{bm}
\usepackage{amsmath}
\usepackage{amsfonts}

\begin{document}

\title{Comparing Effective-One-Body gravitational waveforms to accurate
  numerical data}

\author{Thibault \surname{Damour}}
\affiliation{Institut des Hautes Etudes 
Scientifiques, 91440 Bures-sur-Yvette, France}
\affiliation{ICRANet, 65122 Pescara, Italy} 
\author{Alessandro \surname{Nagar\footnote{Supported by a fellowship from 
        the Istituto Nazionale di Fisica Nucleare (Italy).}}}
\affiliation{Institut des Hautes Etudes 
Scientifiques, 91440 Bures-sur-Yvette, France}
\affiliation{ICRANet, 65122 Pescara, Italy}

\begin{abstract}
  We continue the program of constructing, within the Effective-One-Body (EOB)
  approach, high accuracy, faithful analytic waveforms describing the
  gravitational wave signal emitted by inspiralling and coalescing binary
  black holes. We present the comparable-mass version of a new, {\it resummed}
  3~PN-accurate EOB  quadrupolar waveform that we recently introduced in the
  small-mass-ratio limit. We compare the phase and the amplitude of this
  waveform to the recently published results of a high-accuracy numerical
  simulation of 15 orbits of an inspiralling equal-mass binary black hole
  system performed by the Caltech-Cornell group. We find a remarkable
  agreement, both in phase and in amplitude, between the new EOB waveform and
  the published numerical data. More precisely: (i) in the gravitational wave
  (GW) frequency domain $M\omega <0.08$ where the phase of one of the 
  non-resummed  ``Taylor approximant'' (T4) waveform matches well with the
  numerical relativity one, we find that the EOB phase fares as well,
  while (ii) for higher GW frequencies, $0.08<M\omega\lesssim 0.14$, where the
  Taylor~T4 approximant starts to significantly diverge from the numerical
  relativity phase, we show that the EOB phase continues to match well the
  numerical relativity one. 
  We further propose various methods of tuning the two inspiral flexibility
  parameters, $a_5$ and $v_{\rm pole}$, of the EOB waveform so as
  to ``best fit'' EOB predictions to numerical data.
  We find that the maximal dephasing between EOB and numerical relativity 
  can then be reduced below $10^{-3}$ GW cycles over the entire span (30 GW
  cycles) of the simulation (while, without tuning them, the dephasing 
  is $<8\times 10^{-3}$ cycles). 
  In addition, our resummed EOB amplitude agrees much better with the numerical
  relativity one than any of the previously considered non-resummed,
  post-Newtonian one (including a recently derived, non-resummed 3~PN-accurate one).
  We think that the present work, taken in conjunction with other recent works
  on the EOB-numerical-relativity comparison confirms the ability of the
  EOB formalism (especially in its recently improved avatars) to faithfully
  capture the ``real'' general relativistic waveforms.

\end{abstract}

\date{\today}


\maketitle

\section{Introduction}
\label{sec:intro}
A ground-based network of interferometric gravitational wave (GW) detectors is
currently taking data. Coalescing black hole binaries are among the most
promising GW sources for these detectors. In order to successfully detect
GWs from coalescing black hole binaries and to be able to reliably measure
the source physical parameters, one needs to have in advance a large bank
of  ``{\it templates}'' that accurately represent the GW waveforms emitted
by these binaries. In the terminology of \cite{gr-qc/9708034} one needs 
templates that are both {\it effectual} and {\it faithful}. The construction
of faithful GW templates for coalescing binaries comprising spinning black
holes (with arbitrary masses $m_1$, $m_2$ and spins ${\bf S}_1$, ${\bf S}_2$) 
poses a difficult challenge. Due to the multi-dimensionality of the
corresponding parameter space, it seems impossible for state-of-the-art
numerical simulations to densely sample this parameter space. This motivates
the need to develop {\it analytical} methods for computing (as a function of
the physical parameters $m_1$, $m_2$, ${\bf S}_1$, ${\bf S}_2$) the
corresponding waveforms.
The Effective-One-Body (EOB) 
method~\cite{Buonanno:1998gg,Buonanno:2000ef,Damour:2000we,gr-qc/0103018}
was developed to analytically represent the motion of, and radiation from,
coalescing binary black holes with arbitrary masses and spins. As early
as 2000~\cite{Buonanno:2000ef} this method made several quantitative and
qualitative predictions concerning the dynamics of the coalescence, and the
corresponding waveform, notably: (i) a blurred transition from inspiral to
a ``plunge'' that is just a smooth continuation of the inspiral, (ii) a
sharp transition, around the merger of the black holes, between a continued
inspiral and a ringdown signal, and (iii) estimates of the radiated energy
and of the spin of the final black hole. 

The recent impressive breakthroughs in numerical relativity 
(NR)~\cite{gr-qc/0507014,gr-qc/0602115,Campanelli:2005dd, 
Campanelli:2006gf,Campanelli:2006uy,
Baker:2006yw,Baker:2006vn,Baker:2007fb,Gonzalez:2006md,
arXiv:0706.0740,
Koppitz:2007ev,arXiv:0708.3999,arXiv:0710.3345, Boyle:2007ft,arXiv:0710.0158}
have given us access to extremely valuable, and reliable, information 
about the dynamics and radiation of binary black hole coalescence. 
It is comforting (for theorists) to note that the picture which
is emerging from the recent numerical simulations
(for a review see~\cite{arXiv:0710.1338}) 
broadly confirms the predictions made by the EOB approach. This gives 
us confidence in the soundness of  the various theoretical tools and 
assumptions used in this approach, such as the systematic use of 
{\it resummation} methods, notably Pad\'e approximants 
(as first suggested in~\cite{gr-qc/9708034}).

An important aspect of the EOB approach (which was emphasized early
on~\cite{gr-qc/0103018}) is its {\it flexibility}. As was mentioned in the
latter reference  ``one can modify the basic functions [such as $A(u)$]
determining the EOB dynamics by introducing new parameters corresponding
to (yet) uncalculated higher PN effects.[\dots]. Therefore, when either
higher-accuracy analytical calculations are performed or numerical relativity
becomes able to give physically relevant data about the interaction of
(fast-spinning) black holes, we expect that it will be possible to complete
the current EOB Hamiltonian so as to incorporate this information''. Several
aspects of the EOB flexibility have been investigated early on, such as a 
possible ``fitting'' of a parameter (here denoted as $a_5$), representing 
unknown higher PN effects, to numerical relativity data~\cite{gr-qc/0204011} 
concerning quasi-equilibrium initial configurations~\cite{gr-qc/0106015,gr-qc/0106016},
and the extension of the EOB formalism by several new 
``flexibility parameters''~\cite{gr-qc/0211041}, and notably a parameter, here 
denoted as $v_{\rm pole}$, entering the Pad\'e resummation of the (energy flux
and ) radiation reaction force. 

In view of the recent progress in numerical relativity, the time is ripe 
for tapping the information present in numerical data, 
and for using it to {\it calibrate} the various flexibility parameters of the EOB
approach. This general program has been initiated in a series of recent papers 
which used 3-dimensional numerical relativity 
results~\cite{gr-qc/0610122,arXiv:0704.1964,arXiv:0704.3550,arXiv:0706.3732}. 
In addition, numerical simulations of 
test particles (with an added radiation reaction force) moving
in black hole backgrounds have given an excellent (and well controllable)
``laboratory'' for learning various ways of improving the EOB formalism by
comparing it to numerical data~\cite{arXiv:0705.2519}.
The latter work has introduced a new {\it resummed} 3~PN-accurate quadrupolar
waveform which was shown to exhibit a remarkable agreement with ``exact''
waveforms (in the small mass ratio limit).
In the present paper, we shall present the comparable-mass version of our new,
resummed 3~PN-accurate quadrupolar waveform and compare it to the published 
results~\cite{arXiv:0710.0158} concerning recent high-accuracy numerical simulation
of 15 orbits of an inspiralling equal-mass binary black hole system.
We then show how the agreement between the two (which is quite good even
without any tuning) can be further improved by tuning the two main EOB
flexibility parameters: $a_5$ and $v_{\rm pole}$. Our work will give new
evidence for the remarkable ability of the EOB formalism at describing, in
fine quantitative details, the waveform emitted by a coalescing binary.

\section{Calibrating $v_{\rm pole}$, in the small-mass-ratio case,
from numerical data}
\label{sec:poisson}
As a warm up towards our comparable-mass flexibility study, 
let us first consider the much
simpler small-mass-ratio case, $\nu\ll 1$. Here, $\nu$ denotes the
symmetric mass ratio $\nu=m_1 m_2/(m_1+m_2)^2$ of a binary
system of non-spinning black holes, with masses $m_1$ and $m_2$.
We also denote $M=m_1+m_2$ (``total rest mass''), and $\mu=m_1m_2/M$
(``effective mass for the relative motion''), so that $\nu=\mu/M$.
In the small-mass-ratio limit $\nu\ll 1$, the conservative dynamics
of the small mass (say $m_2\simeq \mu$) around the large one
($m_1\simeq M$) is known, being given by the Hamiltonian describing
a test particle $\mu$ in the background of a Schwarzschild black
hole of mass $M$. On the other hand, the energy flux toward infinity,
say $F=(dE/dt)^{\rm rad}$, or the associated radiation reaction
force ${\cal F}_{\rm RR}$, cannot be analytically computed in closed 
form. One must resort to black hole perturbation theory, whose 
foundations were laid down long ago by Regge and Wheeler~\cite{RW57}, 
and by Zerilli~\cite{Zerilli:1970se} (for the non-spinning case considered 
here). The waveform emitted by a test particle is then computed by 
solving decoupled partial differential equations 
(for each multipolarity $(\ell, m)$ of even or odd parity $\pi$) of the form
\begin{equation}
\label{eq:perturb}
\partial^2_t h^{(\pi)}_{\ell m} -\partial^2_{r_*} h^{(\pi)}_{\ell m} +
V^{(\pi)}_{\ell}(r_{*}) h^{(\pi)}_{\ell m} = S_{\ell m}^{(\pi)} \ ,
\end{equation}
where $V^{(\pi)}_{\ell}$ is an effective radial potential and where
the source term $S^{(\pi)}_{\ell m}$~\cite{Zerilli:1970se,gr-qc/0502028,gr-qc/0502064} 
is linked to the dynamics\footnote{As discussed 
in~\cite{gr-qc/0612096,arXiv:0705.2519} the small-mass-ratio
limit of the EOB formalism leads to a generalization of the
Regge-Wheeler-Zerilli formalism in that the dynamics of the sourcing particle
$\mu$ is not taken to be geodesic, but is assumed to be modified by a
radiation reaction force ${\cal F}_{\rm RR}$. The main issue of interest here
is to optimize the resummation of the analytical approximation to 
${\cal F}_{\rm RR}$, which is given by a badly convergent post-Newtonian expansion
(known only to some finite order).} of $m_2\simeq\mu$ around $m_1\simeq M$.

At this stage we have two options for solving Eq.~(\ref{eq:perturb}):
(i) use numerical methods, or (ii) use an analytical approximation scheme
for solving~(\ref{eq:perturb}) by successive approximations. The numerical
approach led, long ago, to the discovery of several important features of
gravitational radiation in black hole backgrounds, such as the sharp
transition between the plunge signal and a ringing tail when a particle falls
into a black hole~\cite{DRT72}. The analytical approach to
solving Eq.~(\ref{eq:perturb}) by successive approximations, of the
post-Newtonian (PN) type, has been recently driven to unprecedented
heights of sophistication (and iteration order). See~\cite{sasaki_lrr}
for a review.

\begin{figure}[t]   

 \begin{center}
   (a)\includegraphics[width=90 mm, height=77 mm]{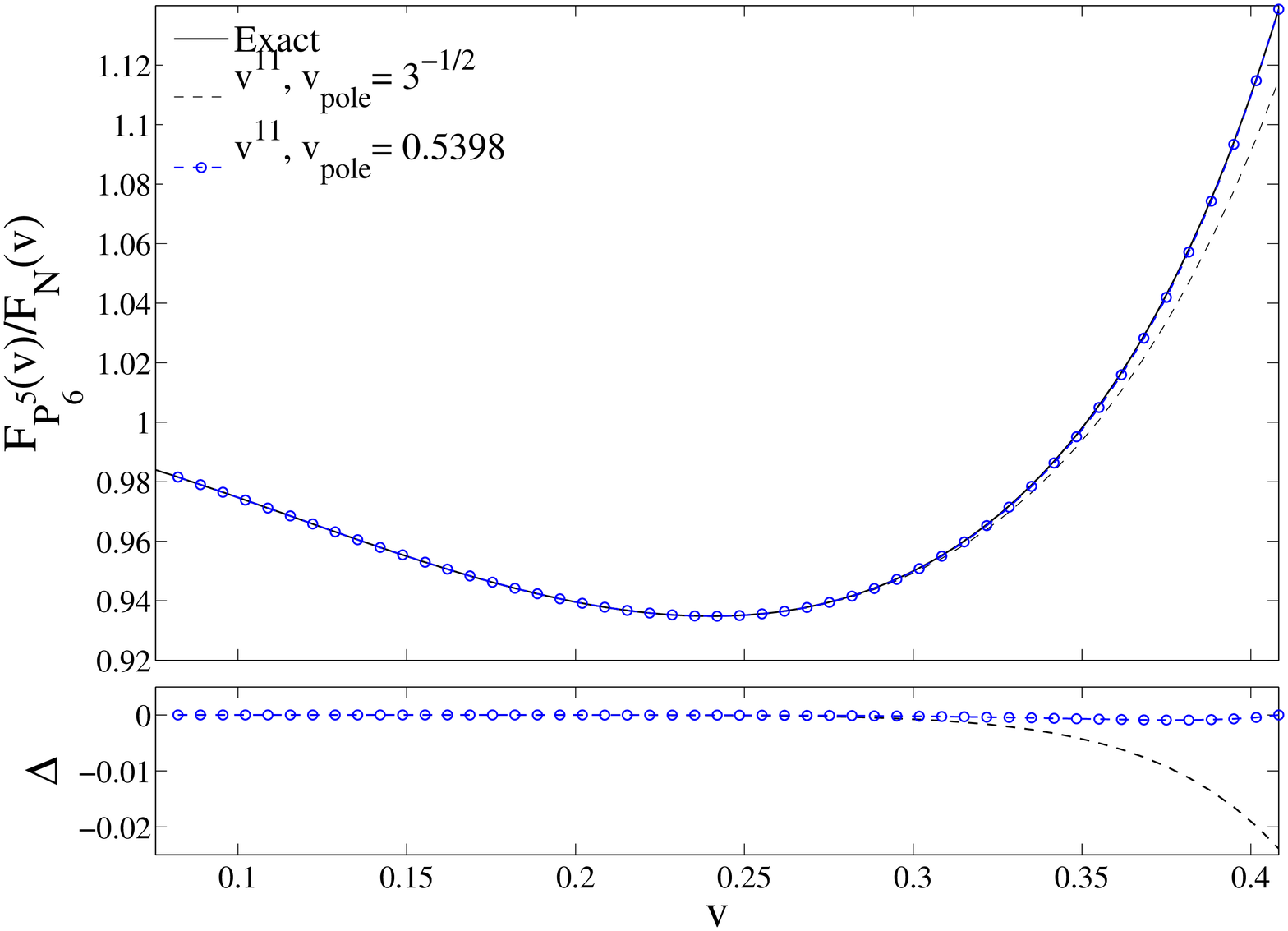}\\
   (b)\includegraphics[width=90 mm, height=77 mm]{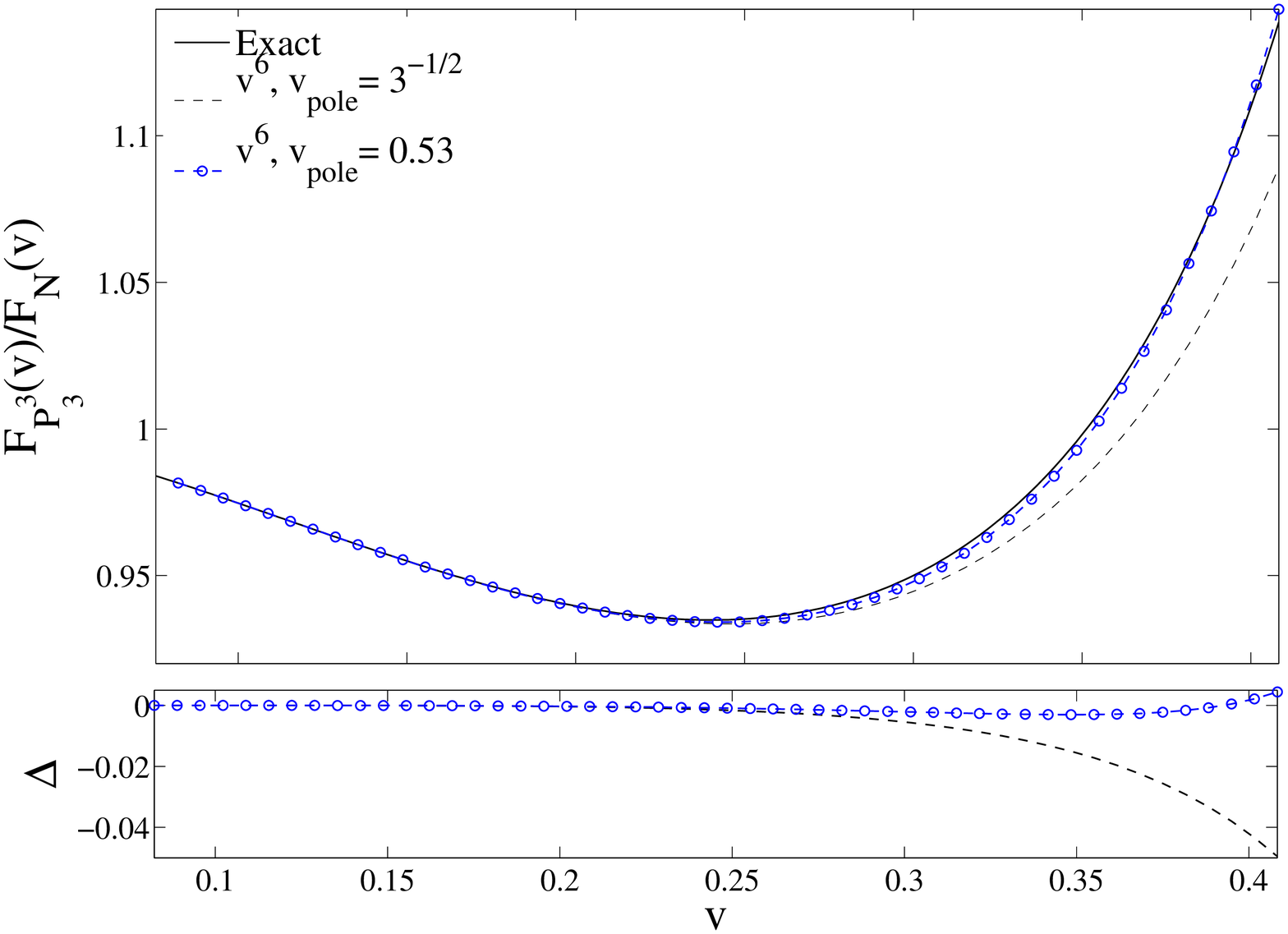}
   \caption{\label{fig:fig1} Panel (a) compares the ``exact''
   Newton-normalized flux function $\hat{F}(v)$~\cite{gr-qc/9505030} 
   to two different Pad\'e resummed, \hbox{$v^{11}$--accurate} analytical flux
   functions: one using the standard value $v_{\rm pole}=1/\sqrt{3}=0.57735$ 
   and the other one using an ``optimized'' flexed value $v_{\rm pole}=0.5398$. The
   bottom part of (a) plots the corresponding 
   differences $\Delta = \hat{F}^{\rm Pade}-\hat{F}^{\rm Exact}$. Panel (b)
   plots the same quantities, except for the fact that it uses only
   $v^6$--accurate analytical flux functions.}
 \end{center}
\end{figure}

Our purpose in this introductory section is to illustrate, on a simple
case, how accurate numerical data can be used to optimize the resummation of
PN-expanded analytical results. We consider the case of a particle on a 
circular orbit. The numerical solution of this problem~\cite{Cutler:1993vq,gr-qc/9505030} 
leads to an accurate knowledge of the radiated energy flux $F$ as a function of the 
orbital radius, or equivalently (and more invariantly) of the ``velocity parameter''
$v=(GM\Omega)^{1/3}$. 
See Fig.~\ref{fig:fig1} where the solid (``Exact'') line represent the
``Newton-normalized flux function''
\begin{equation}
\hat{F}(v)\equiv\dfrac{F(v)}{F_{\rm N}(v)}\,;\quad {\rm with} \quad F_{\rm
  N}(v)\equiv \dfrac{32}{5}\nu^2 v^{10} \ .
\end{equation}
On the other hand, post-Newtonian perturbation theory allows one to compute
$\hat{F}(v)$ as, essentially, a $Taylor$ series in powers of $v$ (modulo the 
appearance of logarithms of $v$ in the coefficients $A_n$ when $n\geq6$,
except for $n=7$)~\cite{sasaki_lrr}, say
\begin{equation}
\label{eq:taylor_flux}
\hat{F}^{\rm Taylor}(v) = 1+ A_2 v^2 + A_3 v^3 + \cdots + A_n v^n + \cdots \ . 
\end{equation}
It was emphasized by Poisson~\cite{gr-qc/9505030} that the successive 
Taylor approximants obtained from Eq.~(\ref{eq:taylor_flux}) converge both 
slowly and erratically to the numerically determined ``exact'' $\hat{F}(v)$. 
Subsequently, Ref.~\cite{gr-qc/9708034}
pointed out that the resummation of the series~(\ref{eq:taylor_flux}) by means
of successive (near diagonal) {\it Pad\'e approximants } led to a much better
sequence of approximants. See Fig.~3 in Ref.~\cite{gr-qc/9708034} for a comparison
between Taylor approximants and Pad\'e approximants. The convergence of the
sequence of Pad\'e approximants was found to be much improved (the $v^5$
approximant being already very close to all its successors), and to be
monotonic. This led to the suggestion of using such Pad\'e approximants also
in the comparable-mass case, though we do not know (yet) the finite-$\nu$ analog
of the exact flux function $F(v;\,\nu)$.

The Pad\'e resummation advocated in Ref.~\cite{gr-qc/9708034} involves one 
{\it  flexibility parameter}, $v_{\rm pole}$, which parametrizes the
location of the (real and positive) pole of the Pad\'e-resummed 
$\hat{F}^{\rm  Pade}(v,\, v_{\rm pole})$ which is closest to the origin
in the complex $v$ plane. Technically speaking, 
$\hat{F}^{\rm Pade}(v;v_{\rm pole})$ is defined as $(1-v/v_{\rm pole})^{-1}$ 
times the relevant near-diagonal Pad\'e approximant~\footnote{Given a certain
order for the Taylor approximant, say $\hat{F}'^{\rm Taylor}=1+\cdots+v^N$, the
general prescription is to resum it with a {\it near-diagonal} Pad\'e,
$P^m_n$, such that $m+n=N$ and $n=m+\epsilon$ with $\epsilon=0$ or $1$. In the
(exceptional) cases where such a near-diagonal Pad\'e contains a ``spurious
pole'' (i.e., a real pole between $0$ and $v_{\rm pole}$), one should use
another choice for $m$ and $n$ (staying as close as possible to the diagonal
$m=n$).} of the $v_{\rm pole}-modified$ Taylor series
$\hat{F}^{'\rm Taylor}(v,v_{\rm pole})\equiv \hat{F}^{\rm
  Taylor}(v)-(v/v_{\rm pole})\hat{F}^{\rm Taylor}(v)=1-v/v_{\rm pole}+A_2 v^2
+\cdots$. Ref.~\cite{gr-qc/9708034} advocated to use, as a fiducial value for
$v_{\rm pole}$, $v_{\rm pole}=1/\sqrt{3}=0.57735$ in the test-mass limit
$\nu\rightarrow 0$, and a slightly larger, $\nu$-dependent value, say 
$v^{\rm  DIS}_{\rm pole}(\nu)$ (motivated by Pad\'e resumming an auxiliary 
``energy function'' $e(v;\,\nu)$ ) given in Eq.~(4.8) there. Here, we point 
out that, when $\nu\rightarrow 0$, a slightly different choice for the
numerical value of $v_{\rm pole}$ can very significantly improve the 
closeness between the Pad\'e flux and the exact (numerical) one.

Our results are displayed in Fig~\ref{fig:fig1} (a) and (b).
In both panels, the solid line represents the ``exact'' result for the
flux function $\hat{F}(v)$ as numerically computed by Poisson. In the upper
part of Fig~\ref{fig:fig1} (a) one compares $\hat{F}^{\rm Exact}(v)$ to two
different Pad\'e ($P^5_6$) approximants resumming the same $v^{11}$-accurate 
(or 5.5 PN) Taylor approximant~\cite{Tagoshi:1994sm}: the ``standard''
$\hat{F}^{\rm Pade}(v; \, v_{\rm pole}=1/\sqrt{3})$ and 
a ``$v_{\rm pole}$-flexed''~\cite{gr-qc/0211041} version of 
$\hat{F}^{\rm Pade}(v;\, v_{\rm  pole})$ using the optimized value
$v_{\rm pole}^{\rm best}(5.5{\rm PN})=0.5398$. This choice of $v_{\rm pole}$ 
yields a Pad\'e approximant which is amazingly close to the exact value.
The lower panel of Fig.~\ref{fig:fig1} (a) exhibits the differences 
$\Delta = \hat{F}^{\rm Pade}-\hat{F}^{\rm Exact}$ for the two choices
of $v_{\rm pole}$. While the standard choice of $v_{\rm pole}$ (namely
$1/\sqrt{3}=0.57735$) leads to a rather good agreement (with $|\Delta|$ being
smaller than $5\times 10^{-3}$ up to $v\simeq 0.355$, which corresponds to
a radius $r=7.93 GM$, and $|\Delta|$ reaching $2.4\times 10^{-2}$ at the
Last Stable Orbit (LSO) at $r=6GM$), the ``flexed choice'' 
$v_{\rm pole}^{\rm best}=0.5398\pm0.0001$ yields an amazing agreement all over
the interval $0\leq v\leq v_{\rm LSO}=1/\sqrt{6}=0.40825$ . The largest  value
of $|\Delta|$ over this interval is ${\rm max}|\Delta|\simeq 9\times 10^{-4}$,
and is reached around $v=0.38$. Note that the 4-digit accuracy quoted for 
$v_{\rm pole}^{\rm best}=0.5398\pm0.0001$ corresponds to (somewhat arbitrarily)
imposing that the value of $|\Delta|$ at the LSO is smaller than about $1\times
10^{-4}$. The rounded off value $v_{\rm pole}=0.54$ would still yield an
amazing fit with ${\rm max}|\Delta|\simeq 10^{-3}$. 

In Fig.~\ref{fig:fig1} (b) we explore what happens when using a much lower
accuracy for the Taylor approximant of the flux. We consider here, as an
example of relevance for the finite $\nu$ case, the case where one starts from
a $v^6$-accurate (3PN) Taylor approximant for the flux\footnote{The result
for the $v^7$-accurate expansion would be very similar and the
final difference would be invisible to the naked eye. However, as we shall
mention below, some problems with spurious poles creep up in the near
diagonal 3.5~PN Pad\'e approximant when $\nu=1/4$ and $v_{\rm pole}\leq 0.55$.
Therefore we prefer to exhibit here the spurious-pole-free 3~PN Pad\'e case.}.
For that case the standard-choice $v_{\rm pole}=1/\sqrt{3}$ still leads 
to a rather good agreement ( with $|\Delta|<10^{-2}$ up to $v\simeq 0.325$ and
$|\Delta|_{\rm LSO}\simeq 5\times 10^{-2}$), while the flexed choice 
$v_{\rm pole}^{\rm best}=0.53$ yields an excellent agreement all over the
interval $0\leq v\leq v_{\rm LSO}$ (with ${\rm max}|\Delta|\simeq 3\times 10^{-3}$
being reached around $v\simeq 0.355$). Though the closeness is less good 
than in the 5.5 PN case ($3\times 10^{-3}$ versus $0.9\times 10^{-3}$), it
is even more amazing to think that, starting from a 3PN-expanded flux function
which (as shown, e.g., in Fig.~3 of Ref.~\cite{gr-qc/9708034}) differs from the exact
result when $0.3\lesssim v\lesssim v_{\rm LSO}$ by about $10\%$, a suitably
flexed Pad\'e resummation can decrease the difference below the $3\times
10^{-3}$ level!

Summarizing: In the small $\nu$ limit, the value of the flexibility parameter
$v_{\rm pole}$ can be calibrated to yield an excellent agreement (from
$3\times 10^{-3}$ to  $0.9\times 10^{-3}$ depending on the PN accuracy) 
between the Pad\'eed flux function $\hat{F}^{\rm Pade}(v;\,
v_{\rm pole})$ and the numerically determined ``exact'' flux 
$\hat{F}^{\rm  Exact}(v)$ all over the interval $0\leq v\leq v_{\rm LSO}$.
This gives an example of the use of accurate numerical data to calibrate
a theoretical flexibility parameter entering the EOB approach. In the
following, we shall consider the equal-mass case, $\nu=1/4$, and investigate
to what extent accurate numerical data~\cite{arXiv:0710.0158} 
can be similarly used to calibrate the two main EOB flexibility parameters $a_5$ 
and $v_{\rm pole}$\footnote{Note that, as already suggested 
in Ref.~\cite{gr-qc/9708034}, one expects the ``true'' value of $v_{\rm pole}$ to
depend on $\nu$. Therefore, we cannot a priori assume that the above best
values, say $v_{\rm pole}^{\rm best}\simeq 0.53$, will yield a close agreement
for the flux function (or the radiation reaction) in the comparable mass case
$\nu\neq 0$.}.

\section{New, resummed 3 PN-accurate EOB inspiral waveform}
\label{sec:waveforms}

After having considered the importance, for fitting high-accuracy numerical
data, of the flexibility parameter $v_{\rm pole}$ in the simpler
small-mass-ratio limit, we wish to move on to the observationally urgent 
comparable mass case $4\nu\sim 1$. As we are going to see, this case involves
{\it two}, rather than one, relevant flexibility parameters: $v_{\rm pole}$
(entering radiation reaction) and $a_5$ (entering the conservative orbital
dynamics). To understand the meaning of these parameters when $4\nu\sim 1$,
let us present the comparable-mass version of the new, improved ``version'' 
of EOB which has been introduced in Ref.~\cite{arXiv:0705.2519} and shown 
there to exhibit a remarkable agreement, in phase and in amplitude, 
with ``exact'' small mass ratio NR waveforms. 
Ref.~\cite{arXiv:0705.2519} considered the small $\nu$ limit, but with the 
clear methodological aim of using this limit to test
improved EOB waveforms defined for any value of $\nu$. We here continue this
program by comparing this improved EOB waveform to the recent numerical
relativity data of~\cite{arXiv:0710.0158}.
The improvements in the EOB approach introduced in 
Ref.~\cite{arXiv:0705.2519} concern several of
the separate ``bricks'' entering this approach. Indeed, it included:
(i) a resummed, 3 PN-accurate description of the inspiral waveform, (ii) a
better description of radiation reaction during the plunge, (iii) a refined
analytical expression of the plunge waveform, and (iv) an improved treatment
of the matching between the plunge and ring-down waveforms. As the present
paper will compare this improved EOB approach to the {\it inspiralling} NR
results of~\cite{arXiv:0710.0158}, we shall only make use here of the improvement (i).

\subsection{Improved, resummed 3PN-accurate waveform}
\label{sbsc:improved}

The new, resummed 3PN-accurate inspiral waveform\footnote{Contrary 
to Ref.~\cite{arXiv:0705.2519} where we used a Zerilli-Moncrief normalized 
waveform $\Psi_{22}$, we use here the same $h_{22}$ normalization as 
Ref.~\cite{arXiv:0710.0614}. They differ simply by a numerical factor: 
$Rh_{\ell  m}=\sqrt{(\ell+2)(\ell+1)\ell(\ell-1)}\left(\Psi_{\ell m}^{(\rm e)} +
{\rm i}\Psi^{(\rm o)}_{\ell m}\right)$.  } 
derived in Ref.~\cite{arXiv:0705.2519} takes the form (when neglecting the 
``non quasi-circular'' flexibility parameters, $a$ and $b$, introduced to
better represent the ``plunge'' which follows the inspiral)
\begin{equation}
\label{eq:EOBh22}
\left(\dfrac{R c^2}{GM}\right) h_{22}^{\rm inspiral}(t)=-8\sqrt{\dfrac{\pi}{5}}\nu
                                        (r_{\omega}\Omega)^2 F_{22} e^{-2{\rm i}\Phi} \ ,
\end{equation}
where $\Phi(t)$ is the EOB orbital phase, $\Omega=\dot{\Phi}$ is the EOB
orbital frequency, $r_{\omega}\equiv r\psi^{1/3}$ is a modified EOB
radius~\footnote{The quantity $r_\omega$ is such that, during adiabatic
inspiral, it is related to $\Omega$ by a standard Kepler-looking law
$\Omega^2r_{\omega}^3=1$, without correcting factors. However, during the
plunge $r_{\omega}$ starts significantly deviating from $\Omega^{-2/3}$~\cite{Damour:2006tr}.}, 
with $\psi$ being defined in Eq.~(22) of Ref.~\cite{Damour:2006tr},
and where the crucial novel PN-improving factor $F_{22}$ is given as the
product of four terms
\begin{equation}
\label{eq:F22}
F_{22}(t) = \hat{H}_{\rm eff} T_{22} f_{22}(x(t)) e^{{\rm i}\delta_{22}(t)} \ .
\end{equation}
Here $\hat{H}_{\rm eff}$ is the {\it effective} EOB Hamiltonian divided by
$\mu$ (it describes the quasi-geodesic dynamics of the ``effective test mass''
$\mu$), and $T_{22}$ is the particularization to $\ell=m=2$ of a resummed
``tail correction factor'' introduced in Ref.~\cite{arXiv:0705.2519}. 
Its explicit expression (in the general, finite $\nu$ case) reads
\begin{equation}
\label{eq:tail}
T_{\ell m} = \dfrac{\Gamma(\ell+1 -2{\rm i}\hat{\hat{k}})}{\Gamma(\ell
  +1)}e^{\pi\hat{\hat{k}}} e^{2{\rm i}\hat{\hat{k}}\log(2 k r_0)} \ ,
\end{equation}
where $\hat{\hat{k}}\equiv G H_{\rm real} m\Omega$ differs from $k=m\Omega$ by
a rescaling involving the {\it real} (rather than {\it effective})
EOB Hamiltonian. This ``tail factor'' is the exact resummation of an 
infinite number of ``leading logarithms'' appearing in the perturbative
multipolar-post-Minkowskian (MPM) 
expansion~\cite{Blanchet:1985sp,Blanchet:1986dk,Blanchet:1989ki,Damour:1990gj,Damour:1990ji}
of ``tail effects'' in the $(\ell, m)$ radiative moment. For instance, at the
leading order in the monopole$\times$multipole interaction the radiative
quadrupole $U_{ij}(T_R)$ contains a tail integral~\cite{Blanchet:1992br} 
\begin{equation}
2GI\int_{0}^{\infty}d\tau M^{(4)}_{ij}(T_R-\tau)
\left[\log \left(\dfrac{\tau}{2r_0}\right)+\dfrac{11}{12}\right] \ ,
\end{equation}
while at the next to leading order it contains a tail 
integral~\cite{gr-qc/9710038}
\begin{align}
2G^2 I^2\int_0^\infty d\tau
M_{ij}^{(5)}(T_R-\tau)
\bigg[&\log^2\left(\dfrac{\tau}{2r_0}\right) \nonumber \\
      &+c_1\log\left(\dfrac{\tau}{2r_0}\right)+c_0\bigg] \ .
\end{align}
[Here, $I$ denotes the monopole of the source, 
i.e. \hbox{$I=M_{\rm ADM}=H_{\rm real}$}]. 
The factor $T_{\ell m}$ resums the infinite series of 
the contributions to $U_{\ell m}$ proportional to
\begin{equation}
G^n I^n \int_0^\infty d\tau M^{(\ell+1+n)}_{\ell m}
(\tau_R-\tau)\log^n\left(\dfrac{\tau}{2r_0}\right) \ .
\end{equation} 

The real factor $f_{22}(x)$ was computed in Ref.~\cite{arXiv:0705.2519} (as
indicated in footnote 8 there) to 3~PN accuracy for all values of $\nu$ by
starting from the 3~PN-accurate multipolar post-Minkowskian results 
of Refs.~\cite{Blanchet:1995ez,gr-qc/0105098,gr-qc/0406012,Blanchet:2005tk,Blanchet:2001ax}.
The explicit form of its (PN) ``Taylor'' expansion reads
\begin{align}
\label{eq:f22}
f_{22}^{\rm Taylor}(x) &= 1 + \dfrac{1}{42}\left(-86 + 55\nu\right) x
\nonumber \\
&+\dfrac{1}{1512}\left(-4288 - 6745\nu + 2047\nu^2\right) x^2 \nonumber \\
&+\bigg(
          \dfrac{21428357}{727650}   - \dfrac{856}{105}{\rm eulerlog}(x) -
	  \dfrac{34625}{3696}  \nu   \nonumber\\
&       + \quad\dfrac{41}{96}\pi^2\nu - \dfrac{227875}{33264}\nu^2 
        + \dfrac{114635}{99792}\nu^3 \bigg) x^3 \nonumber \\
&       + \bigg( -\dfrac{5391582359}{198648450} + \dfrac{36808}{2205}{\rm eulerlog}(x)\bigg) x^4 \nonumber \\
&       + \bigg( -\dfrac{93684531406}{893918025}+\dfrac{458816}{19845}  {\rm
	  eulerlog}(x)\bigg)x^5 \nonumber \\
&       + {\cal O}(\nu x^4) + {\cal O}(x^6) \ .   
\end{align}
where ${\rm eulerlog}(x)\equiv \gamma_{\rm E} + 2\log2 +\dfrac{1}{2}\log x$.
For greater accuracy, we have added in Eq.~(\ref{eq:f22}) the small $\nu$
limit of the 4PN and 5PN contributions (as deduced from the results 
of~\cite{Tagoshi:1994sm,Tanaka:1997dj}).

Finally, the additional phase $\delta_{22}(t)$ is given by
\begin{equation}
\label{eq:delta22}
\delta_{22} = \dfrac{7}{3}H_{\rm real}\Omega + \dfrac{428}{105}\pi
\left(H_{\rm  real}\Omega\right)^2 -24\nu x^{5/2} \ .
\end{equation}
At the time of the writing of~\cite{arXiv:0705.2519} we had derived the full
$\nu$-dependent waveform (\ref{eq:EOBh22})-(\ref{eq:delta22}), except for the
accurate value of the coefficient of $\nu x^{5/2}$ in Eq.~(\ref{eq:delta22}),
because this value was not meaningful for us as it could be absorbed in the
``non quasi-circular phase flexibility parameter'' $b$ included in Eq.~(14) 
of~\cite{arXiv:0705.2519}. Taking $b=0$, we have recently derived from scratch
the value $-24$ for this coefficient\footnote{As was emphasized in 
Refs.~\cite{arXiv:0706.0726,arXiv:0710.0158,arXiv:0710.0614} the small
additional phase term $\propto \nu x^{5/2}$ has anyway very little effect on
observable quantities during the inspiral.}.
In the meantime, Kidder~\cite{arXiv:0710.0614} has independently realized
that it might be useful to compute the $\ell =m=2$ part of the waveform to
3~PN accuracy and has derived a 3~PN-accurate non-resummed $(2,2)$ 
waveform [See also~\cite{Berti:2007fi} for an earlier 
2.5~PN-accurate derivation of the (non-resummed) $(2,2)$ waveform]. 
We have compared our result (\ref{eq:F22})-(\ref{eq:delta22})
with his and found perfect agreement (when PN-reexpanding our result).

Speaking of choices, there are more to be made to convert the PN-expanded
amplitude factor $f_{22}^{\rm Taylor}$, Eq.~(\ref{eq:f22}), into a better,
``resummed'' EOB waveform factor. As said in Ref.~\cite{arXiv:0705.2519}, we
propose to improve the convergence properties of the Taylor expansion
(\ref{eq:f22}) by replacing it by a suitable Pad\'e approximant. We use
the upper diagonal $(3,2)$ Pad\'e, i.e. we use in our calculations 
$f_{22}(x;\,\nu)=P^3_2\left[f_{22}^{\rm Taylor}(x ;\,\nu)\right]$\footnote{A 
technical remark concerning the Pad\'e approximants we use (both
for the flux function $F(v)$ and the waveform $f_{22}(x)$): contrary to the
prescription suggested in Ref.~\cite{gr-qc/9708034}, we find simpler (and numerically
more or less equivalent) not to factor out the log-dependent terms (appearing
at 3~PN and beyond), but instead to define the Pad\'e approximants by
considering the logarithms appearing in the PN expansion on par with the
normal numerical Taylor coefficients: e.g., 
$1+c_1x+\cdots +(c^0_3 + c^1_3\log x)x^3$ is Pad\'eed by first Padeing the
usual Taylor series $1+c_1 x+\cdots + c_3 x^3$, and then replacing
$c_3\rightarrow c^0_3 + c^1_3\log x$ in the result.}. 
The final choice we need to make concern the argument $x(t)$ of 
$f_{22}(x(t))$ in Eq.~(\ref{eq:f22}). As discussed in 
Ref.~\cite{arXiv:0705.2519} this argument is ``degenerate'' during the inspiral
in that it can be equivalently expressed in various ways in terms of the
dynamical variables of the system, namely (in the general finite $\nu$ case)
$x=\Omega^{2/3}=(r_{\omega}\Omega)^2=1/r_{\omega}$. We emphasized that some
choices might be better than others to automatically capture some non
quasi-circular effects during the plunge. However, as we are here concerned
with the inspiral phase, we do not expect that the precise choice of $x(t)$
will matter. Some preliminary checks indicate that this is indeed the
case. For simplicity, we shall use here the argument $x(t)=\Omega^{2/3}$ 
for $f_{22}(x(t))$ in Eq.~(\ref{eq:F22}).

\subsection{Effective One Body relative dynamics}
\label{sbsc:eob_dynamics}

Let us briefly recall here the EOB construction of the relative dynamics 
of a two-body system (for more details on the recent improvements in the 
EOB approach see~\cite{arXiv:0704.3550,arXiv:0705.2519}). 
The EOB approach to the general relativistic two-body dynamics is a 
{\it non-perturbatively resummed} analytic technique which has been 
developed in 
Refs.~\cite{Buonanno:1998gg, Buonanno:2000ef,Damour:2000we,gr-qc/0103018,Damour:2006tr,arXiv:0704.3550,arXiv:0705.2519}.
The EOB approach uses as basic input the results of PN and MPM 
perturbation theory, and then ``packages'' this PN-expanded 
information in special {\it resummed} forms, which are expected 
to extend the validity of the PN results beyond their normal 
weak-field-slow-velocity regime into (part of) the strong-field-fast-motion
regime. At the practical level, and for what concerns the part of the EOB
approach which deals with the relative orbital dynamics, the method consists
of two fundamental ingredients: (i) the ``real Hamiltonian'' $H_{\rm real}$,
and (ii) the radiation reaction force ${\cal F}_\varphi$.

It is convenient to replace the adimensionalized radial momentum $p_r$
(conjugate to the EOB adimensionalized radial coordinate $r=R/M$) by the
conjugate $p_{r_*}$ to the ``EOB tortoise radial coordinate''
\begin{equation}
\label{eq:eob_tortoise}
\dfrac{dr_*}{dr}=\left(\dfrac{B}{A}\right)^{1/2} \; ; \quad B\equiv
\dfrac{D}{A} \ .
\end{equation}
In terms of $p_{r_*}$ (and after the rescaling 
$\hat{H}_{\rm eff}\equiv H_{\rm  eff}/\mu$, 
$p_{\varphi}\equiv P_{\varphi}/(\mu M)$) the real, 3~PN-accurate 
Hamiltonian~\cite{Damour:2000we} reads
\begin{equation}
\label{eq:Hreal}
H_{\rm real}(r,p_{r_*},p_{\varphi})\equiv 
                       \mu \hat{H}_{\rm real}=M\sqrt{1+2\nu\left(\hat{H}_{\rm eff}-1\right)} \ ,
\end{equation}
with 
\begin{equation}
\label{eq:Heff}
\hat{H}_{\rm eff}(r, p_{r_*},p_{\varphi})
\equiv \sqrt{ p_{r_*}^2 +A(r)\left(1+\dfrac{p_{\varphi}^2}{r^2}+z_3\dfrac{p_{r_*}^4}{r^2}\right)} \ ,
\end{equation}
where $z_3=2\nu\left(4-3\nu\right)$, and where the PN expansion of the crucial
radial potential $A(r)\;\left(\equiv -g_{00}^{\rm effective}\right)$ has the 
form~\cite{Buonanno:1998gg,Damour:2000we} 
\begin{align}
\label{eq:ATaylor}
A^{\rm Taylor}(u) &= 1-2u + 2\nu u^3 \nonumber \\
                  &+\left(\dfrac{94}{3}-\dfrac{41}{32}\pi^2\right)\nu u^4 + a_5\nu u^5 + {\cal O}(\nu u^6) \ ,
\end{align}
with $u=1/r$. As suggested in Ref.~\cite{gr-qc/0103018} we have parametrized
the presence of presently uncalculated 4~PN (and higher) contributions to 
$A(u)$ by adding a term $+a_5(\nu)u^5$, with the simple form
$a_5(\nu)=a_5\nu$. [Indeed, it was remarkably found, both at the 1~PN, 
2~PN~\cite{Buonanno:1998gg}, and the 3~PN~\cite{Damour:2000we} levels, 
that, after surprising cancellations between higher powers of $\nu$ in 
the various contributions to the coefficient 
$a_n(\nu) = a^1_n\nu + a^2_n\nu^2 + \cdots$ of $u^n$  in $A(u)$,
only the term linear in $\nu$ remained for $n$=2,\footnote{%
                Actually, for $n=2$, i.e. the 1~PN contribution 
		to $A(u)$, the cancellations even led to a complete 
		cancellation $a_2(\nu)=0$!}
3 and 4. Ref.~\cite{gr-qc/0103018} introduced this term with the idea that
``one might introduce a 4~PN contribution $+a_5(\nu)u^5$ to $A(u)$, as a free
parameter in constructing a bank of templates, and wait until LIGO-VIRGO-GEO
get high signal-to-noise ratio observations of massive coalescing binaries to
determine its numerical value''. We do not dispose yet of such real
observations, but, as substitutes we can (as started in
Refs.~\cite{gr-qc/0204011,arXiv:0706.3732}) try to use numerical simulations 
to determine (or at least constrain) the value of the unknown parameter 
$a_5$. This is what we shall do below, where we shall compare our results 
to previous ones.

As discussed in~\cite{Damour:2000we}, the most robust choice\footnote{And the
simplest one ensuring continuity with the $\nu\to 0$ limit.} for resumming the
Taylor-expanded function $A(u)$ is to replace it by the following Pad\'e
approximant $A(u)\equiv P^1_4[A^{\rm Taylor}(u)]$. Similarly, the secondary
metric function $D(u)=P^0_3[D^{\rm Taylor}(u)]$ where $D^{\rm Taylor}(u)$ is 
given in Eq.~(2.19) of Ref.~\cite{gr-qc/0103018}.

The EOB equations of motion for $(r,\,r_*,\, p_{r_*},\,\varphi,\, p_{\varphi})$
are then explicitly given by Eqs.~(6-11) of Ref.~\cite{arXiv:0704.3550}. They
are simply Hamilton's equations following from the
Hamiltonian~(\ref{eq:Hreal}), except for the $p_{\varphi}$ equation of motion
which reads
\begin{equation}
\label{eq:RR}
\dfrac{dp_{\varphi}}{dt}=\hat{\cal F}_{\varphi} \ ,
\end{equation}
where, following Refs.~\cite{gr-qc/9708034,Buonanno:2000ef} the r.h.s.
contains a {\it resummed} radiation reaction force, which we shall 
take in a form recently suggested in Ref.~\cite{Damour:2006tr}, 
namely
\begin{equation}
\label{eq:flux}
{\hat{\cal F}}_{\varphi}\equiv \dfrac{{\cal F}_{\varphi}}{\mu}=-\dfrac{32}{5}\nu
\Omega^5 r_{\omega}^4 \dfrac{f_{\rm DIS}(v_{\varphi};\,\nu,\,v_{\rm
      pole})}{1-v_{\varphi}/v_{\rm pole}} \ ,
\end{equation}
where $v_{\varphi}\equiv \Omega r_{\omega}$, $r_{\omega}\equiv r\psi^{1/3}$
(with $\psi$ defined as in Eq.~(22) of Ref.~\cite{Damour:2006tr}). Here
$f_{\rm DIS}$ denotes a suitable Pad\'e resummation of the quantity denoted 
$\hat{F}'^{\rm Taylor}(v,v_{\rm pole})$ above, i.e., the Taylor expansion of 
$(1-v/v_{\rm pole})\hat{F}^{\rm Taylor}(v;\, \nu)$ where $\hat{F}^{\rm Taylor}$
is the Newton-normalized (energy or) angular momentum flux along circular
orbits (expressed in terms of $v_{\rm circ}=\Omega^{1/3}$ for comparable-mass
circular orbits). Here again, to completely define $\hat{\cal F}_{\varphi}$ we
must clearly state what is the starting Taylor-expanded result that we use,
and how we resum it. For greater accuracy, we are starting from 
\begin{align}
\hat{F}^{\rm Taylor}(v;\,\nu)&=1+ A_2(\nu) v^2 + A_3 (\nu) v^3 + A_4(\nu) v^4\nonumber \\
                             &+A_5(\nu) v^5 + A_6(\nu, \log v) v^6 + A_7(\nu) v^7 \nonumber \\ 
                             &+ A_8(\nu=0, \log v) v^8
\ ,
\end{align}
where we have added to the known 
3.5~PN-accurate~\cite{Blanchet:1995ez,gr-qc/0105098,gr-qc/0406012,Blanchet:2005tk,Blanchet:2001ax} 
comparable-mass flux the small-mass-ratio 4~PN contribution~\cite{Tagoshi:1994sm}. 
Then, we use as Pad\'e approximant of this (quasi-)$v^8$-accurate expansion 
$f_{\rm DIS}(v;\, v,\, v_{\rm pole})\equiv P^4_4\left[(1-v/v_{\rm
    pole})\hat{F}^{\rm Taylor}(v;\,\nu)\right]$. We indeed found that this specific
(diagonal) Pad\'e approximant (as well as the less accurate $P^{3}_{3}$ one)
was robust under rather large variations of the numerical value of 
$v_{\rm pole}$ (by contrast to other ones, such as $P^3_4$ or $P^4_3$, 
which exhibit spurious poles when $v_{\rm pole}$ becomes too small).
Finally, note that, for integrating the EOB dynamics from some finite starting
radius (or frequency) we need some appropriate initial conditions. 
Refs.~\cite{Buonanno:2000ef,gr-qc/0211041,gr-qc/0508067} indicated how to
define some ``post-adiabatic'' initial conditions. In view of the high
accuracy of the NR data of~\cite{arXiv:0710.0158} (and notably of their
extremely reduced eccentricity), we found useful to go beyond 
Refs.~\cite{Buonanno:2000ef,gr-qc/0211041,gr-qc/0508067} and to define some,
iterated  ``post-post-adiabatic'' initial data allowing us to start
integration at a radius $r=15$.

Summarizing: In the comparable-mass case, the phasing, and the amplitude, 
of the new, resummed 3~PN-accurate inspiral waveform introduced in 
Ref~\cite{arXiv:0705.2519} is given by inserting the solution of the EOB
dynamics (given by Eqs.~(\ref{eq:Hreal})-(\ref{eq:flux})) into the 
waveform~(\ref{eq:EOBh22})-(\ref{eq:delta22}). This waveform depends
on two {\it flexibility parameters}, $a_5$ and $v_{\rm pole}$, that
parametrize (in an effective manner) current uncertainties in the EOB
approach: $a_5$ parametrizes uncalculated 4~PN and higher {\it orbital
effects}, while $v_{\rm pole}$ parametrizes uncertainties in the 
{\it resummation of radiation effects} (also linked to $\nu$ dependent 
4~PN and higher radiative effects).

\section{Comparing the new, resummed EOB waveform to accurate numerical data}
\label{sec:boyle}

Thanks to the recent breakthroughs in numerical relativity one can now start 
to make detailed comparisons between EOB waveforms and numerical relativity
ones. Working with high-accuracy data can further allow us to calibrate 
the ``flexibility parameters''~\cite{gr-qc/0211041} entering extended versions 
of the EOB formalism. A first step in this direction was recently taken by 
Buonanno et al.~\cite{arXiv:0706.3732} by utilizing numerical gravitational 
waveforms generated in the merger of
comparable-mass binary black holes. However, the merger data that were used
were relatively short (about 7 inspiralling orbits before merger when
$\nu=1/4$). Here, we shall instead consider the information contained in a
recent, high accuracy, low-eccentricity simulation covering 15 orbits of an
inspiralling equal-mass binary black hole~\cite{arXiv:0710.0158}. [See also the previous
results of the Jena group which covered $\sim 9$ inspiralling 
orbits~\cite{arXiv:0706.1305}]. 

Boyle et~al.~\cite{arXiv:0710.0158} have published their results in the form 
of {\it differences} between the numerical relativity data and various, 
Taylor-type PN predictions
We recall that there are many ways of defining some Taylor-type PN 
waveforms. Ref.~\cite{gr-qc/0010009} introduced a nomenclature which 
included three sorts of PN-based ``Taylor approximants''; from Taylor~T1 to 
Taylor~T3, as well as several other resummed approximants (Pad\'e, and EOB). 
In addition, the definition of each such Taylor approximant makes two 
further choices: the choice of PN accuracy on the phasing, and the choice 
of PN accuracy in the amplitude. A fourth Taylor approximant, T4, was 
also considered in previous NR-PN 
comparison~\cite{gr-qc/0610122,gr-qc/0612024,arXiv:0704.1964}
and Ref.~\cite{gr-qc/0610122} had pointed out that it seemed to yield a phasing 
close to the NR one. Boyle et al.~\cite{arXiv:0710.0158} confirmed 
this ``experimental'' fact and found that Taylor T4 at 3.5 PN phasing 
accuracy agreed much better with their long, accurate simulations than 
the other Taylor approximants.

Here we shall use as EOB quadrupolar metric waveform the new, resummed
3~PN-accurate inspiral waveform explicitly presented above, 
say $h_{22}^{\rm EOB}(t; \, a_5, \, v_{\rm pole})$, defined in Eq.~(\ref{eq:EOBh22}) 
and the following equations of the previous section. 
Note, however, that~\cite{arXiv:0710.0158} uses as basic 
gravitational radiation variable the $\ell=m=2$ projection of the 
corresponding Weyl curvature quantity, which is related to the {\it metric}
waveform $h_{22}$ by
\begin{equation}
\label{eq:psi4}
\left(\dfrac{R c^2}{GM}\Psi_4^{22{\rm X}}\right)(t)\equiv\dfrac{\partial^2}{\partial
  t^2}\left(\dfrac{Rc^2}{GM}h_{22}^{\rm X}\right)(t)\equiv A_{\rm X}(t) e^{-{\rm i}\phi_{\rm X}(t)} \ .
\end{equation}
Here, $A$ and $\phi$ denote the {\it amplitude} and {\it phase} of the
curvature wave considered by Boyle et al.~\cite{arXiv:0710.0158}. We have
introduced a label X which will take, for us, {\it three}  values: 
X$\equiv$NR denotes the numerical relativity result of
Ref.~\cite{arXiv:0710.0158}, X$\equiv$EOB denotes the improved EOB waveform
presented in the previous section and X=T4 will denote the ``Taylor T4''
waveform highlighted in Ref.~\cite{arXiv:0710.0158}
(as well as in previous PN-NR comparisons~\cite{arXiv:0704.1964,gr-qc/0612024}) 
as giving a particularly good fit. 

The precise definition of this T4 waveform (as used in
Ref.~\cite{arXiv:0710.0158}) is as follows. As indicated in
Eq.~(\ref{eq:psi4}), $\Psi_4^{22{\rm T4}}$ is obtained by taking two time
derivatives of a corresponding metric T4 waveform $h_{22}^{\rm T4}$. The latter
waveform is defined (if we understand correctly the combined statements of 
Refs.~\cite{arXiv:0710.0158} and ~\cite{arXiv:0710.0614}) by the following procedure.
First, one defines a certain ``T4 orbital phase'' $\Phi_{\rm T4}(t)$ by
integrating the ODEs
\begin{align}
\label{eq:T41}
\dfrac{d\Phi_{\rm T4}}{dt} &=\dfrac{x^{3/2}}{GM} \ , \\
\label{eq:T42}
\dfrac{dx}{dt}             &=\dfrac{64\nu}{5GM}x^5a_{3.5}^{\rm Taylor}(x) \ ,
\end{align}
where 
\begin{align}
\label{eq:3.5Taylor}
a_{3.5}^{\rm Taylor}(x) &= 1 + \bar{a}_2(\nu) x + \bar{a}_3(\nu) x^{3/2} +
\bar{a}_4(\nu) x^2 \nonumber \\
            &+ \bar{a}_5(\nu) x^{5/2} + \bar{a}_6(\nu,\log x)x^3 + \bar{a}_7(\nu) x^{7/2} \ ,
\end{align}
is the 3.5 PN Taylor approximant (for finite $\nu$) to the Newton-normalized
ratio (flux-function)/(derivative of energy function)=$F(v)/E'(v)$,
which enters the adiabatic evolution of the orbital phase (see,
e.g.~\cite{gr-qc/9708034,gr-qc/0010009}). In the relevant case $\nu=1/4$, 
$a_{3.5}^{\rm Taylor}(x)$ is explicitly given as the quantity within curly
braces on the r.h.s. of Eq.~(45) of~\cite{arXiv:0710.0158}. Having in hand the
result, $\Phi_{\rm T4}(t)$, 
$\Omega_{\rm T4}(t)\equiv d\Phi_{\rm T4}/dt\equiv x^{3/2}_{\rm T4}/GM$, of
integrating Eqs.~(\ref{eq:T41})-(\ref{eq:T42}) one then defines a
nPN-accurate, T4 $(\ell=2,m=2)$ waveform by truncating to order $x^{n/2}_{\rm T4}$
included  the Taylor series
\begin{align}
\label{eq:h22Kidder}
&\left(\dfrac{c^2 R}{GM}h_{22}^{\rm T4}\right)(t)=-8\sqrt{\dfrac{\pi}{5} }\nu
e^{-2{\rm i}\Phi_{\rm T4}(t)}x_{\rm T4}\nonumber\\
&\times \bigg[1+\tilde{h}_2 x_{\rm T4} +
  \tilde{h}_3 x_{\rm T4}^{3/2}+\tilde{h}_4x^2_{\rm T4}+\tilde{h}_5x^{5/2}_{\rm
    T4} + \tilde{h}_6 x^3_{\rm T4}\bigg] \ ,
\end{align}
where the coefficients $\tilde{h}_n$ are obtained from the coefficients $h_n$ 
in the Taylor-expanded 3~PN-accurate (2,2) waveform derived 
by Kidder~\cite{arXiv:0710.0614} 
by {\it setting to zero} all the terms proportional 
to $\log(x/x_0)$ or $\log^2(x/x_0)$ (but keeping the 
separate $\log x$ term entering $h_6$), and replacing 
$\nu=1/4$. 

Note that our resummed waveform~(\ref{eq:EOBh22})-(\ref{eq:delta22}) 
differs from the 3~PN-accurate version of ~(\ref{eq:h22Kidder}) in several
ways: (i) the ``orbital phase'' evolution is given, for us, by the EOB  
resummed dynamics, (ii) we do not neglect the phase terms linked to 
$\log(x/x_0)$, but include them either in the resummed tail factor 
$T_{22}$ or in $\delta_{22}$, and (iii) we resum the amplitude of 
$h_{22}$ by factoring both $\hat{H}_{\rm eff}$ and the modulus of $T_{22}$, 
and by Padeing $f_{22}$.

Ref.~\cite{arXiv:0710.0158} presents in their Fig.~19 the differences, 
in phase and amplitude (of the radiative Weyl-curvature components $\Psi_4$), 
between Taylor~T4~3.5/2.5 (i.e., 3.5~PN in phase and 2.5 PN in amplitude) 
and the (unpublished) corresponding Caltech-Cornell NR data, say
\begin{align}
\label{eq:diffT4_phase}
(\Delta\phi)_{\rm T4NR} &\equiv \phi_{\rm T4}(t)-\phi_{\rm NR}(t) \ , \\
\label{eq:diffT4_mod}
\left(\dfrac{\Delta A}{A}\right)_{\rm T4NR}&\equiv \dfrac{A_{\rm T4}(t)-A_{\rm NR}(t)}{A_{\rm NR}(t)} \ ,
\end{align} 
as plotted in the two panels of Fig.~19 of Ref.~\cite{arXiv:0710.0158}. 
Our main aim here is to compare the 
``numerical data''~(\ref{eq:diffT4_phase})-(\ref{eq:diffT4_mod}) 
to the corresponding theoretical predictions made by the EOB formalism, say
\begin{align}
\label{eq:diffEOB}
(\Delta\phi)_{\rm T4EOB}&\equiv \phi_{\rm T4}(t)-\phi_{\rm EOB}(t) \ , \\
\left(\dfrac{\Delta A}{A}\right)_{\rm T4EOB}&\equiv \dfrac{A_{\rm T4}(t)-A_{\rm EOB}(t)}{A_{\rm EOB}(t)} \ .
\end{align} 
As just said, as the most complete, and best plotted data, concern 
Taylor~T4~3.5/2.5, we shall use the 3.5~PN accurate Eq.~(\ref{eq:3.5Taylor})
but only consider the 2.5~PN truncation of the Taylor-expanded 
waveform~(\ref{eq:h22Kidder}) (i.e. a waveform essentially 
contained in the 2.5~PN $h_+$ and $h_\times$ results of Arun et al.~\cite{gr-qc/0404085}). 
[As we use the T4 waveform only as an {\it intermediary} between the NR 
and EOB results, we are allowed to use any convenient ``go between'', 
even if its PN accuracy differs from the (formal) one of our resummed EOB
waveform]. 

To effect the comparison between NR and EOB, i.e., to compute the crucial 
difference $\phi_{\rm EOB}-\phi_{\rm NR}$, we needed 
to extract actual numerical data from Fig.~19 of~\cite{arXiv:0710.0158}. 
We did that in several ways. First, we measured (with millimetric accuracy; 
on an A3-size version of the left panel of Fig.~19) sufficiently many points 
on the solid upper curve (Taylor~T4~3.5/2.5 matched
at $M\omega_4\equiv 0.1$ )
\footnote{Ref.~\cite{arXiv:0710.0158} computes various differences 
           $\Delta^{\omega_m}\phi(t) = \phi^{\omega_m}_{\rm  T4}(t'_{\rm T4})-\phi_{\rm NR}(t)$
          where, given a ``matching'' frequency $\omega_m$, $\phi_{\rm  T4}^{\omega_m}(t'_{\rm T4})$ 
          denotes a version of $\phi_{\rm T4}(t_{\rm T4})$ which is shifted in $\phi$
          and in $t$ so that $0=\Delta^{\omega_m}\phi(t)=d\Delta^{\omega_m}\phi(t)/dt$
          at the moment $t^{\omega_m}_{\rm NR}$ where $d\phi_{\rm NR}/dt=\omega_m$.
          We shall denote the four matching frequencies used in~\cite{arXiv:0710.0158} as
          $M\omega_1\equiv 0.04$, $M\omega_2 = 0.05$, $M\omega_3=0.063$, $M\omega_4 = 0.1$}
to be able to replot (after ``splining'' our measured points) this upper curve
with good (visual) accuracy. 
We could then use this splined version of eleven (adequately distributed) 
selected points on the upper
curve on the left panel of Fig.~19 as our basic (approximate) ``numerical
data''. It gives us a (continuous) approximation to the $\omega_4$-matched
phase difference 
$\Delta^{\omega_4}\phi_{\rm T4NR}(\bar{t}^{\omega_4})
=\phi^{\omega_4}_{\rm T4}(\bar{t}'_{\rm T4})
-\phi_{NR}(\bar{t}^{\omega_4})$. [Here, $\bar{t}^{\omega_4}$ denotes the
time shifted so that $\bar{t}^{\omega_4}=0$ corresponds to the $\omega_4$
matching point between T4 and NR]. As we have separately computed 
$\phi_{\rm  T4}(t_{\rm T4})$ by integrating Eqs.~(\ref{eq:T41})-(\ref{eq:T42})
(and that it is easy to shift it to obtain $\phi^{\omega_4}_{\rm
  T4}(\bar{t}'_{\rm T4}))$ we have thereby obtained an approximation to
$\phi_{\rm NR}(\bar{t}^{\omega_4})$. We then shift again the time argument to
our basic EOB dynamical time so as to obtain $\phi_{\rm NR}(t_{\rm EOB})$
(with the condition that $\omega_{\rm EOB}(t^{\omega_4}_{\rm EOB})=\omega_4$
corresponds to $\bar{t}^{\omega_4}=0$). 

There are then several ways of comparing $\phi_{\rm NR}(t_{\rm EOB})$ to the
EOB phasing $\phi_{\rm EOB}(t_{\rm EOB})$, obtained from the procedure explicated above.
We wish to emphasize here that there is a useful way of dealing with the
information contained in (any) phasing function $\phi_{\rm X}(t_{\rm X})$
where X=NR, EOB, T4, etc. Indeed, a technical problem concerning any such
phasing function is the presence of two shift ambiguities: a possible
arbitrary shift in $\phi_{\rm X}$ ($\phi_{\rm X}\to \phi_{\rm X}+c_{\rm X}$)
and a possible arbitrary shift in the time variable $t_{\rm X}$ 
($t_{\rm X}\to t_{\rm X}+\tau_{\rm X}$). Similarly to what one does in
Euclidean plane geometry where one can replace the Cartesian equation of a
curve $y=y(x)$ by its {\it intrinsic} equation $K=K(s)$ (where $K$ is the
curvature and $s$ the proper length), we can here (in presence of a different
symmetry group) replace the shift-dependent phasing 
function $\phi_{\rm X}(t_{\rm X})$ by the shift-independent {\it intrinsic} 
phase evolution equation: 
$d\omega_{\rm X}/dt_{\rm X}=\alpha_{\rm  X}(\omega_{\rm X})$, where 
$\omega_{\rm X}\equiv d\phi_{\rm X}/dt_{\rm X}$ (for simplicity we shall use here 
$M=m_1+m_2=1$). It is also convenient to factor out of the {\it phase acceleration}
$\alpha_{\rm X}(\omega)$ its ``Newton'' approximation\footnote{Note 
 that we are dealing here with the gravitational (curvature)
 wave frequency $\omega$. If we were dealing with the orbital frequency
 $\Omega=d\Phi/dt$, we would instead consider the following reduced
 phase acceleration $(d\Omega/dt)/(C_\nu\Omega^{11/3})=A_{\Omega}(\Omega)$
 with $C_\nu=(96/5)\nu$. },
\begin{align}
\label{eq:alpha_omega}
\alpha_N(\omega)\equiv c_\nu\omega^{11/3}; 
\qquad 
c_\nu=\dfrac{12}{5}2^{1/3}\nu 
\end{align}
and to consider the reduced phase acceleration function $a_{\omega}^{\rm X}$
defined by
\begin{equation}
\label{eq:aomega}
\dfrac{\dot{\omega}_{\rm X}}{c_\nu\omega_{\rm X}^{11/3}} 
                     = a_{\omega}^{\rm X}(\omega_{\rm X}) \ .
\end{equation}
\begin{figure}[t]   
 \begin{center}
   \includegraphics[width=90 mm, height=72 mm]{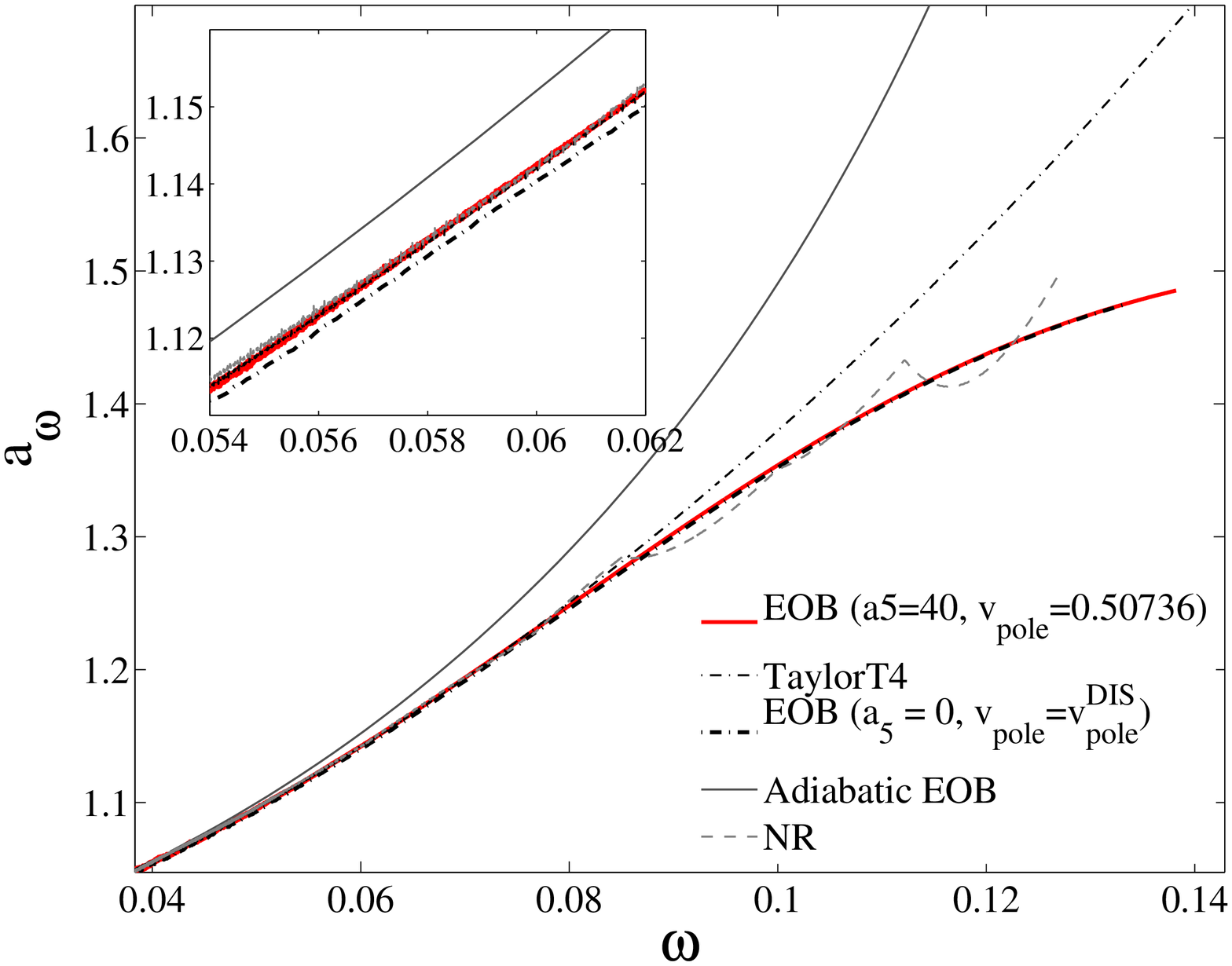}
    \caption{ \label{fig:fig2} Reduced phase-acceleration curves as defined
    in Eq.~(\ref{eq:aomega}) (with $M\equiv m_1+m_2=1$). The inset highlights 
    how the ``tuned'' EOB curve nearly coincides (for $\omega\lesssim 0.08$) 
    with the NR (and T4) curves, while the ``non-tuned'' EOB one lies slightly below.}
 \end{center}
\end{figure}

Independently of the label X the function $a_{\omega}^{\rm X}\to 1$ as $\omega \to 0$.
The phasing comparisons then boil down to comparing the various (reduced)
phase acceleration functions $a_{\omega}^{NR}(\omega)$, $a_{\rm \omega}^{\rm T4}(\omega)$
and $a_{\omega}^{\rm EOB}(\omega;\,a_5,v_{\rm pole})$.
The last two functions can be straightforwardly (numerically) computed from
the formulas written above. As for $a_{\omega}^{\rm NR}(\omega)$ it is, in
principle, also straightfiorwardly computable from 
$\phi_{\rm NR}(t_{\rm  NR})$. 
However, as we do not have access to an accurate estimate of $\phi_{\rm NR}(t_{\rm  NR})$
but only to a rather rough cubic spline approximation to it, we can only compute an
even rougher estimate of $a_{\omega}^{NR}(\omega)$. [As is well known, taking
(two!) derivatives of approximate results considerably degrades the
accuracy.] Still, as we think that this is the conceptually clearest way of
presenting the comparison, we used the data we had in hand to compute the
various ``acceleration curves'' presented in Fig.~\ref{fig:fig2}. 

Actually, it is instructive to include further phase-accelerations in the
comparison. In Fig.~\ref{fig:fig2} we show the following phase-acceleration 
functions  (versus $\omega$, i.e., $M\omega$):
(i) Taylor T4 3.5/2.5, (ii) NR, (iii) a standard ``non-tuned'' EOB with
$a_5 =0$ (i.e. essentially the 3~PN approximation) and 
$v_{\rm pole}=v^{\rm DIS}_{\rm pole}(\nu)$~\cite{gr-qc/9708034}
for $\nu=1/4$, which corresponds to our current knowledge, (iv) a ``tuned'' EOB with
$a_5=40$ and $v_{\rm pole}=0.5074$ (see below), and finally (v) 
the {\it  adiabatic} EOB for $a_5=40$ and $v_{\rm pole}=0.5074$.
Here, the {\it adiabatic} approximation to $a_{\omega}$ is that defined by the
usual adiabatic approach to inspiral phasing (see e.g.~\cite{gr-qc/9708034}),
leading to $a_{\omega}(\omega)=\hat{F}(v)/\hat{E}'(v)$ where
$v=(\omega/2)^{1/3}$ and where $\hat{F}$ is the Newton-normalized circular
flux and $\hat{E}'$ the Newton-normalized derivative of the circular energy
function. [When applying these general concepts to the EOB we need, as
discussed in Ref.~\cite{Buonanno:2000ef}, to use the analytical, adiabatic
approximation to EOB inspiral, with, notably, $p_{\varphi}=j^{\rm adiabatic}(u)$
obtained by solving $\partial H_{\rm EOB}/\partial r=0$ with $p_r=0$].

Several preliminary conclusions can be read off Fig.~\ref{fig:fig2}.

(i) In the frequency domain (say $M\omega<0.08$) where the T4~3.5 phasing 
matches well with the NR phasing, both the standard ``non-tuned'' EOB 
($a_5=0$, $v_{\rm pole}=v^{\rm DIS}_{\rm pole}$) and ``tuned'' EOB phasing (defined above) 
match well with the NR phasing. However, a closer look at the acceleration
curves (see inset) shows that the ``tuned'' EOB phasing agrees better with NR
(and T4). [The ``non-tuned'' $a_{\omega}^{\rm EOB}$ is slightly below
  $a_{\omega}^{\rm NR}$ (and $a_{\omega}^{\rm T4})$ by roughly 
  $1.5\times 10^{-3}$ when $M\omega\sim 0.06$ ].

(ii) For higher frequencies ($0.08<M\omega\lesssim 0.14$), Taylor T4
3.5/2.5 starts to significantly diverge from the NR phasing.
\footnote{Though Ref.~\cite{arXiv:0710.0158} tends to mainly emphasize how
          well Taylor~T4~3.5/2.5
          agrees with the NR phasing one should note that the high curvature
          of the upper $\omega_4$ curves when $M\omega\gtrsim 0.08$ in the
          left panel of Fig.~19, and the subsequent fast rise of all the 
	  $\Delta^{\omega_m}\phi$, are clear signals that Taylor~T4~3.5/2.5
          starts to significantly ({\it and increasingly}) diverge from the
          NR phasing.}
By contrast, both the standard ``non-tuned'' EOB phasing and the 
``tuned'' EOB one continue to match quite well the NR phasing. 
This will be shown below by using other diagnostics than the acceleration
curves. Indeed, when $M\omega \gtrsim 0.08$ our ``NR acceleration curve''
exhibits fake oscillations which come from our use of a coarse approximation 
to NR data. The visible ``kinks'' in our NR acceleration curve are due to our 
taking (numerical) second derivative of a {\it cubic spline} interpolant of 
{\it approximate} NR data points. 
We expect that the exact ``NR acceleration curve'' (computed with accurate 
numerical data instead of our approximate ones) will be
a smooth curve lying close to the two EOB curves in Fig.~\ref{fig:fig2}.

(iii) The fact that the {\it adiabatic} EOB curve diverges quite early, and
upwards, from the full EOB curve is a confirmation of the conclusion derived 
in Ref.~\cite{Buonanno:2000ef} (see Figs.~4 and~5 there), namely that , 
`` in the equal mass case $\nu=1/4$ the adiabatic approximation starts to
significantly deviate from the exact evolution quite before one reaches the
LSO''. This further confirms the suggestion of~\cite{arXiv:0710.0158} that the
good early ($M\omega<0.08$) agreement between T4 and NR is coincidental.

Because of our lack of an accurate knowledge of $a_{\omega}^{\rm NR}$, we cannot 
use the acceleration curves of Fig.~\ref{fig:fig2} to make any accurate comparison
between EOB and NR data. In the following we shall use other tools for doing
this comparison and, in particular, for constraining the values of $a_5$ and
$v_{\rm pole}$.

The first tool we shall use consists in selecting among our eleven
approximate points on the $\Delta^{\omega_4}\phi_{\rm T4NR}$ curve two special 
ones, namely
\begin{align}
\label{eq:deltabwd}
\Delta^{\omega_4}\phi_{\rm T4NR}(t^{\omega_4}_{\rm NR}-1809M)&\equiv 
                 \delta_4^{\rm bwd}\simeq 0.055 \ , \\
\label{eq:deltafwd}
\Delta^{\omega_4}\phi_{\rm T4NR}(t^{\omega_4}_{\rm NR}+44.12M)&\equiv
\delta_4^{\rm fwd}\simeq 0.01 \ ,
\end{align}
to which we shall refer as the (main) ``backward'' and ``forward'' $\omega_4$
data. In addition, we also measured a couple of selected points on the $\omega_2$- and
$\omega_3$-matched lower $\Delta\phi$ curves. Namely,
\begin{align}
\Delta^{\omega_2}\phi(t_{\rm NR}^{\omega_2}+1000M)&\equiv \delta_2\simeq-0.01 \ ,\\
\label{eq:delta3}
\Delta^{\omega_3}\phi(t^{\omega_3}_{\rm NR}-1000M&\equiv\delta_3\simeq
-4.3\times 10^{-3} \ .
\end{align}
We can then use, in a numerically convenient way, these data to quantitatively
compare (with an hopefully reasonable numerical accuracy) NR to EOB by 
considering four {\it ratios}, $\rho_{\omega_2}$,
$\rho_{\omega_3}$, $\rho_{\omega_4}^{\rm bwd}$, $\rho_{\omega_4}^{\rm fwd}$
(where we recall that $\omega_2 = 0.05$, $\omega_3=0.063$ and $\omega_4=0.1$), with
\begin{equation}
\label{eq:ratio_rho}
\rho_{\omega_m}(a_5, v_{\rm pole})\equiv\dfrac{\Delta^{\omega_m}
               \phi_{\rm T4EOB}\left( t^{\omega_m}_{\rm NR}+\delta
               t_m\right)}{\delta_m} \ ,
\end{equation}
and $\omega_m=\omega_2$, $\omega_3$ and $\omega_4^{\rm bwd}$ or $\omega_4^{\rm fwd}$.

\begin{figure}[t]   
 \begin{center}
   \includegraphics[width=90 mm, height=72 mm]{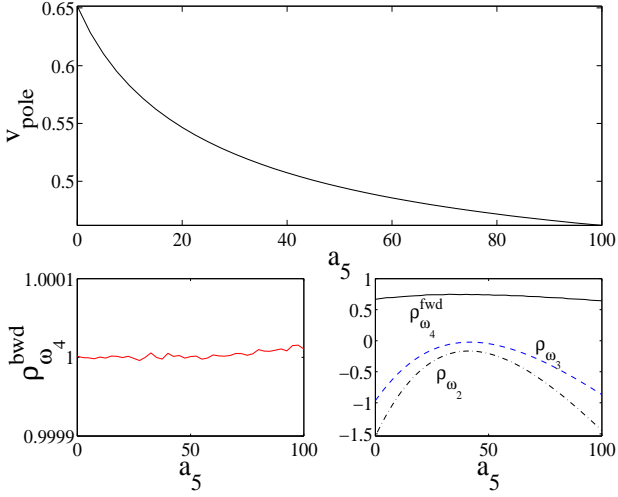}
    \caption{ \label{fig:fig3} Correlation between $v_{\rm pole}$ and $a_5$
    (top panel) obtained by imposing the constraint~(\ref{eq:ratio4}). The
    numerical accuracy with which Eq.~(\ref{eq:ratio4}) is satisfied is
    displayed in the left-bottom panel. The right-bottom panel displays the
    extent to which, as $a_5$ varies, the other ratios $\rho_{\omega_m}$,
    Eq.~(\ref{eq:ratio_rho}), approximate unity.}
 \end{center}
\end{figure}

If our approximate measures (given in Eqs.~(\ref{eq:deltabwd})-(\ref{eq:delta3})) 
of the various $\delta_m$'s were accurate, a perfect match between NR and EOB
would correspond to having all those ratios equal to unity:
$\rho_{\omega_2}(a_5, v_{\rm pole})=1$, 
\hbox{$\rho_{\omega_3}(a_5, v_{\rm  pole})=1$}, 
$\rho_{\omega_4}^{\rm bwd}(a_5, v_{\rm pole})=1$, and
\hbox{$\rho_{\omega_4}^{\rm fwd}(a_5, v_{\rm pole})=1$}.
This would give four equations for two unknowns ($a_5$ and $v_{\rm pole}$). Even
if we had exact values for the various $\delta_m$'s, we do not, however, expect
that there would exist special values of $a_5$ and $v_{\rm pole}$ for which
{\it all} these ratios would be equal to one. Indeed, $a_5$ and $v_{\rm pole}$ are 
only ``effective'' parameters that are intended to approximately mimic an
infinite number of higher $\nu$-dependent, resummed PN-effects. The best we
can hope for is to find values of $a_5$ and $v_{\rm pole}$ allowing one to
give a good overall match between $\phi_{\rm NR}(t)$ and $\phi_{\rm EOB}(t)$ (or
$a_{\omega}^{\rm NR}$ and $a_{\omega}^{\rm EOB}$). To investigate this issue,
it is then convenient to focus first on only {\it one} comparison observable.
We choose $\rho_{\omega_4}^{\rm bwd}$ because it is, among the data which we
could measure with reasonable accuracy, the one which has the largest ``lever
arm''. [Indeed, it corresponds to some weighted integral of the difference
$a_{\omega}^{\rm EOB}-a_{\omega}^{\rm NR}$ over a significantly extended
frequency interval]. Imposing the constraint
\begin{equation}
\label{eq:ratio4}
\rho^{\rm bwd}_{\omega_4}(a_5, v_{\rm pole})=1 \ ,
\end{equation}
then gives a precise way of exploring which extended EOB models best match the
NR phasing . Note first that this equation could have no solutions. [For
instance, if we were using the {\it adiabatic} approximation to EOB there
would be no solutions]. To admit solutions is already a sign that EOB can
provide a much better match to NR than T4. Then, the solutions could exist only if
both $a_5$ and $v_{\rm pole}$ are close to some ``preferred'' values. 
Actually, we found that Eq.~(\ref{eq:ratio4}) defines a 
{\it continuous curve} in the $(a_5, v_{\rm pole})$
plane\footnote{Consistently 
               with what was found for lower approximations, and for
               the presently computable contributions to $a_5$~\cite{gr-qc/0211041},
	       we expect that $a_5\geq 0$, and we shall therefore only work in
               the corresponding half plane.}.

\begin{figure}[t]   
 \begin{center}
   \includegraphics[width=90 mm, height=72 mm]{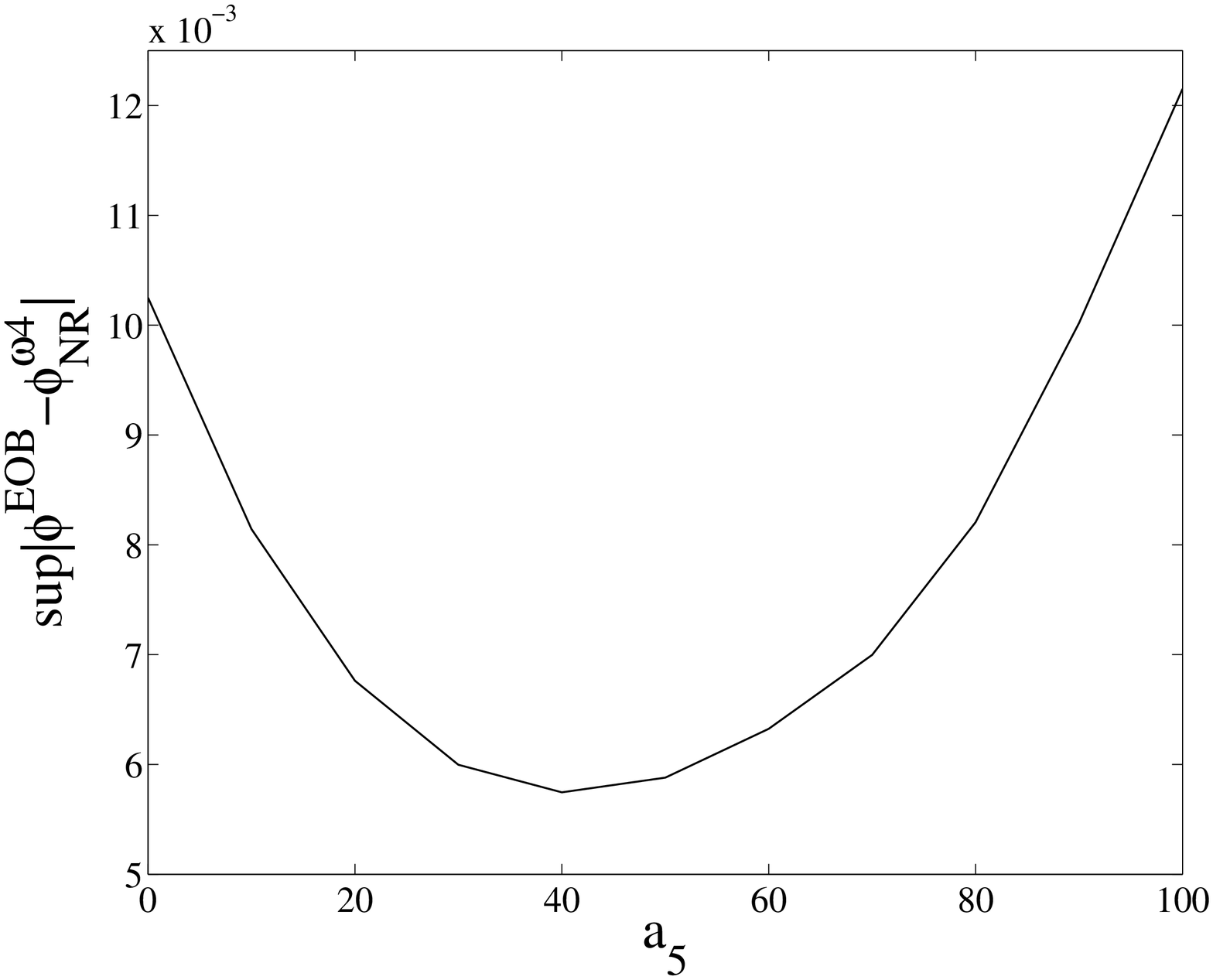}
    \caption{ \label{fig:fig4} The $L_{\infty}$ norm of the phase difference
    between EOB (when $v_{\rm pole}$ is correlated to $a_5$ as in
    Fig.~\ref{fig:fig3}) and numerical relativity, as defined by Eq.~(\ref{eq:linfinity}).}
 \end{center}
\end{figure}

For all values of $a_5\geq 0$, we (numerically) found a unique value of
$v_{\rm pole}$ satisfying the constraint (\ref{eq:ratio4}). This continuous
curve is plotted in the upper panel of Fig.~\ref{fig:fig3}. When remembering
that Eq.~(\ref{eq:deltabwd}) is only 
approximate,\footnote{We estimate the accuracy of our measurement result 
Eq.~(\ref{eq:deltabwd}) to be such that the
\hbox{``backward time-shift''}, corresponding to a r.h.s. exactly equal 
to $0.055$, is $(-1809\pm 15)M$.} 
we have to mentally replace the continuous curve in the upper panel 
of Fig.~\ref{fig:fig3} by a narrow valley of ``best fitting'' values of 
$(a_5, v_{\rm pole})$. Let us first remark that this valley extends only on a
rather small range of values of $v_{\rm pole}$, around 0.55. It is comforting
that this range includes the values that were previously suggested for $v_{\rm pole}$:
namely $v^{\rm usual}_{\rm pole}(\nu=0)=1/\sqrt{3}=0.57735$, 
$v_{\rm pole}^{\rm DIS}(\nu=1/4)\simeq 0.6907$,
$v_{\rm pole}^{\rm best}(\nu=0)\simeq 0.54$ (discussed above).

To go beyond this result and see whether the other measurements constrain the
value of $a_5$, we plot on the lower, right panel of Fig.~\ref{fig:fig3} the values
of the ratios $\rho_{\omega_2}$, $\rho_{\omega_3}$ and $\rho_{\omega_4}^{\rm fwd}$
{\it along} the $\rho_{\omega_4}^{\rm bwd}=1$ curve. As, along this curve,
$v_{\rm pole}$ is a function of $a_5$, the above three ratios depend only on $a_5$.
Ideally, we would like to find values of $a_5$ for which the remaining ratios
are all close to unity. [Given the coarse nature of our measurements, 
we cannot expect to get exactly unity]. We see on Fig.~\ref{fig:fig3} that the ratio 
$\rho_{\omega_4}^{\rm fwd}$ is reasonably close to unity for most values of
$a_5$. By contrast, the two other ratios $\rho_{\omega_2}$ and $\rho_{\omega_3}$
happen to have the {\it wrong sign}. This negative sign means, in terms of the
phase-acceleration curves of Fig.~\ref{fig:fig2}, that around frequencies
$\omega_2$ and $\omega_3$, $a_{\omega}^{\rm NR}(\omega)$ is slightly 
{\it  above} $a_{\omega}^{\rm T4}(\omega)$, while it seems that 
$a_{\omega}^{\rm  EOB}(\omega)$ tends to be generally slightly 
{\it below} $a_{\omega}^{\rm T4}(\omega)$. On the other hand, for larger
frequencies, it seems clear that $a_{\omega}^{\rm NR}$ crosses 
$a_{\omega}^{\rm  T4}$ to become {\it below} $a_{\omega}^{\rm T4}$, and to
become in rather good agreement with $a_{\omega}^{\rm EOB}$. At this stage,
the best we can do is to say that an overall best match between EOB and NR
will be obtained when $a_5$ belongs to a rather large interval 
(say $10\lesssim a_5\lesssim 80$) centered around $a_5\simeq 40$, where  
$\rho_{\omega_2}$ and $\rho_{\omega_3}$ are negative, but rather small
(say $-0.5\lesssim \rho_{\omega_3}\lesssim 0$)

\begin{figure}[t]   
 \begin{center}
   \includegraphics[width=90 mm, height=77 mm]{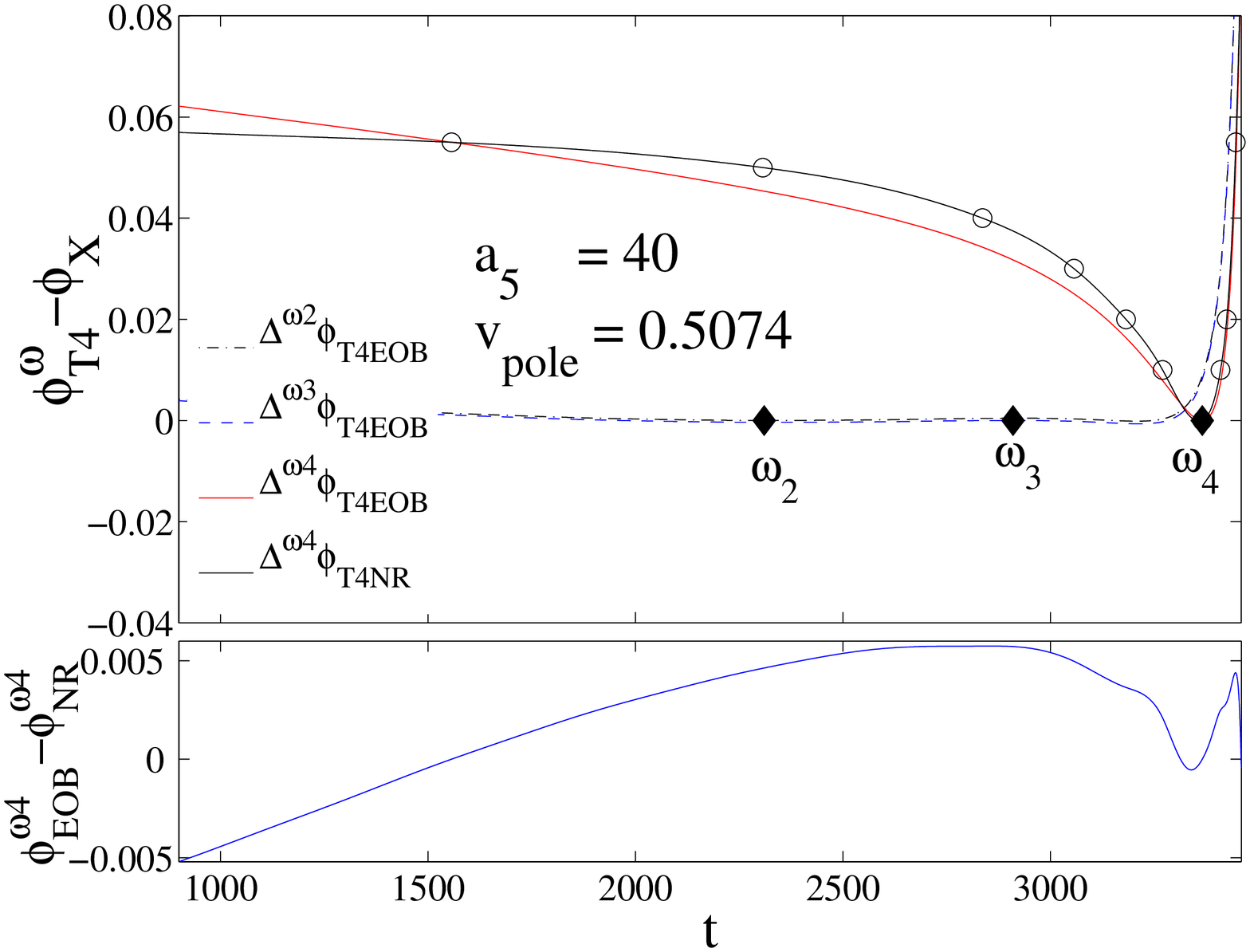}
    \caption{\label{fig:fig6} The upper panel compares various phase differences 
     $\Delta^{\omega_m}\phi_{\rm T4X}$ versus time (with $M=1$), $\omega_m$ 
     denoting a matching frequency and the label X being either EOB or NR. 
     The lower panel exhibits the $\omega_4$--matched phase difference 
     between EOB and NR. The flexibility parameters of EOB have been tuned 
     here to $a_5=40$ and $v_{\rm pole}=0.5074$.}
 \end{center}
\end{figure}

To get another, potentially better measure of the ``closeness'' between NR and
EOB we looked at the ``$L_{\infty}$'' distance between the two functions
$\phi_{\rm NR}(t)$ and $\phi_{\rm EOB}(t)$ on the time interval (in EOB time)
$900M\leq t_{\rm EOB}\leq 3460M$ (which roughly corresponds to the time
interval plotted in Fig.~19 of Ref.~\cite{arXiv:0710.0158}).
More precisely, we computed the quantity
\begin{equation}
\label{eq:linfinity}
L_{\infty}(a_5) \equiv \sup_{900M\leq t_{\rm EOB}\leq 3460M}
                \left|\phi_{\rm EOB}(t_{\rm EOB})-\phi_{\rm
		  NR}^{\omega_4}\left(t'^{\omega_4}_{\rm NR}\right)\right| \ ,
\end{equation}
where $\phi_{\rm NR}^{\omega_4}\left(t'^{\omega_4}_{\rm NR}\right)$ is matched
to the EOB phase at $\omega_4$, and where the EOB was constrained to lie along
the curve $v_{\rm pole}(a_5)$ plotted in Fig.~\ref{fig:fig3} 
(i.e., satisfying Eq.~(\ref{eq:ratio4})). We show this $L_{\infty}$ norm 
in Fig.~\ref{fig:fig4}. This Figure displays the remarkable agreement between 
EOB and NR phasing over an interval where T4 exhibits a clear dephasing with 
respect to NR. Indeed, Fig.~19 of~\cite{arXiv:0710.0158} shows that on this 
interval all 
Taylor~T4~3.5 templates dephase by $\approx 0.08$ radians (because of the 
divergence at the end, corresponding to the divergence of the acceleration 
curves in Fig.~\ref{fig:fig2} when $\omega\gtrsim 0.08$). By contrast, 
the dephasing between EOB and NR can be as small as 0.006 radians if 
$30\lesssim a_5\lesssim 52$, or 0.008 radians if $10\lesssim a_5\lesssim 80$. 
Again, we find that a largish interval of $a_5$ values centered around 
$a_5\sim 40$ seems to be preferred (when $v_{\rm pole}$ is correlated 
to $a_5$ via the curve of Fig.~\ref{fig:fig3}) to give the best possible
overall match between EOB and NR.

\begin{figure}[t]   
 \begin{center}
   \includegraphics[width=90 mm, height=77 mm]{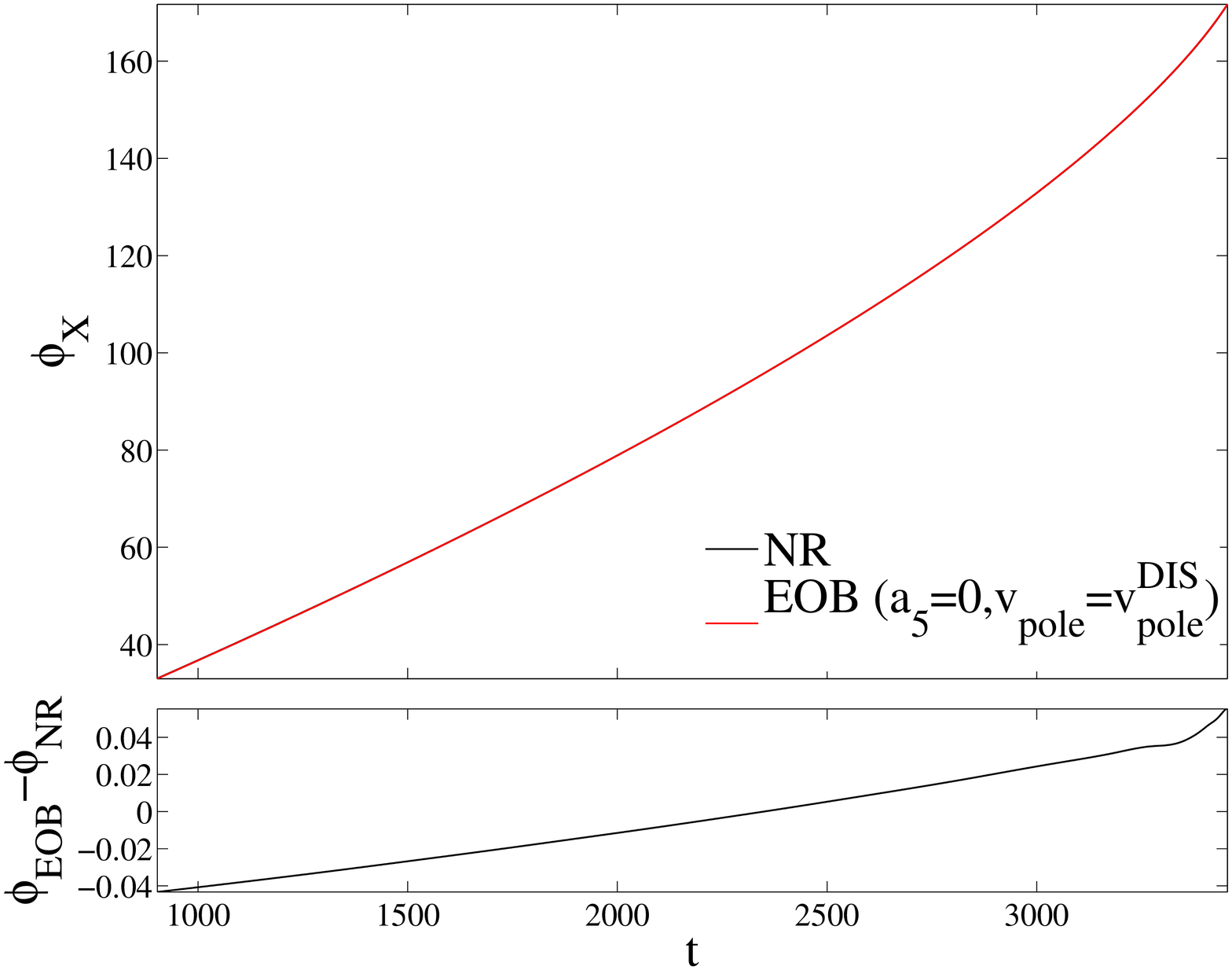}
    \caption{ \label{fig:fig5} Comparison between the standard, ``non-tuned'' 
    EOB ($a_5=0$, $v_{\rm pole}=v_{\rm pole}^{\rm DIS}(\nu=1/4)=0.6907)$ and NR. The top panel shows that
    the gravitational wave phases $\phi_{\rm EOB}$ and $\phi_{\rm NR}$ 
    (versus time) are nearly indistinguishable to the naked eye. The 
     bottom panel quantifies the small difference between the two.}
 \end{center}
\end{figure}

To give a better feeling of how well EOB matches NR phasing all over the time
interval explored by the simulation of Ref.~\cite{arXiv:0710.0158}, we plot 
in Fig.~\ref{fig:fig6} the superposition of the upper curve in Fig.~19 
of~\cite{arXiv:0710.0158} (i.e., the difference $\Delta^{\omega_4}\phi_{\rm
  T4NR}$, as measured and splined by us) with the corresponding EOB difference 
$\Delta^{\omega_3}\phi_{\rm T4EOB}$, for the values 
$a_5=40$, $v_{\rm  pole}=0.5074$ approximately corresponding to the smallest 
$L_{\infty}$ norm in Fig.~\ref{fig:fig4}. We also plot the $\omega_2$- and
$\omega_3$-matched phase differences $\Delta^{\omega_2}\phi_{\rm T4EOB}$
and $\Delta^{\omega_3}\phi_{\rm T4EOB}$. Apart from the slightly wrong
curvatures of the $\omega_2$- and $\omega_3$- curves (for $\omega\lesssim
0.08$), this Figure exhibits a truly remarkable visual agreement with the left
panel of Fig.~19 of~\cite{arXiv:0710.0158}. It exhibits again two facts: (i)
the EOB phasing agrees extremely well with the NR one on the full time
interval ($900M\leq t_{\rm EOB}\leq 3460M$), (ii) by contrast 
Taylor~T4~3.5/2.5 starts diverging from EOB when $\omega\gtrsim 0.08$ 
in precisely the same way that it diverges from NR. 
In the bottom panel of Fig.~\ref{fig:fig6} we give a precise quantitative 
measure of the difference between EOB and NR  phasings by plotting the 
$\omega_4$-matched difference 
$\phi_{\rm EOB}(t)-\phi_{\rm NR}^{\omega_4}\left(t'_{\omega_4}\right)$. 
This phase difference vanishes both when 
$\omega_{\rm EOB}\left(t^{\omega_4}_{\rm EOB}\right)=\omega_4$ (by construction),
and at the time $t_{\rm EOB}^{\rm bwd}=t_{\rm EOB}^{\omega_4}-1809M$ (by our
optimized choice of the link $v_{\rm pole}=v_{\rm pole}(a_5)$, such that 
Eq.~(\ref{eq:ratio4}) holds). We see how, indeed (in agreement with
Fig.~\ref{fig:fig4}) the dephasing remains smaller, in absolute value, than
about 0.006 radians, i.e. 0.001 GW cycles.

This remarkably small dephasing concerns a ``tuned'' EOB phasing (with
optimized flexibility parameters $a_5$ and $v_{\rm pole}$). However, as it is
clear on Fig.~\ref{fig:fig2}, even the standard , ``non-tuned'' EOB phasing
corresponding to our current analytical knowledge $a_5=0$, 
$v_{\rm  pole}=v_{\rm pole}^{\rm DIS}(\nu)$, agrees quite well with the NR 
phasing over the entire simulation time. To exhibit this important fact in
quantitative detail we compare in Fig.~\ref{fig:fig5} the (splined) NR phase 
$\phi_{\rm NR}(t')$ (after suitable shifts in $\phi$ and $t$) to the standard, 
``non-tuned'' EOB phase $\phi_{\rm EOB}(t)$ 
($a_5=0$, $v_{\rm pole}=v_{\rm  pole}^{\rm DIS}(\nu)$). As the {\it visual}
agreement (top panel) is too good to allow one to distinguish the two
curves, we show (bottom panel) the phase difference 
$\phi_{\rm EOB}(t)-\phi_{\rm  NR}(t')$. As expected, the dephasing is less good
than in the above ``tuned'' case, but it remains impressively good: $\pm 0.05$
radians, i.e., $\pm 0.008$ GW cycles, over the full time interval 
$900M\leq t_{\rm EOB}\leq 3460M$.

\begin{figure}[t]   
 \begin{center}
   \includegraphics[width=90 mm, height=77 mm]{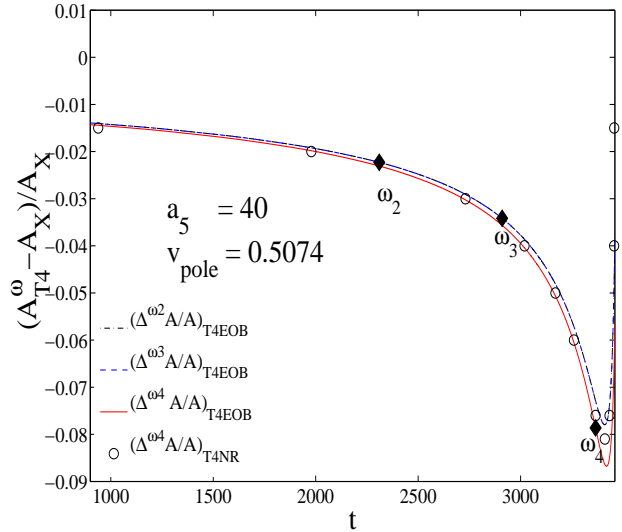}
    \caption{ \label{fig:fig7} Comparison between relative amplitude
    differences $(\Delta^{\omega_m}A/A)_{\rm T4X}$ versus time, $\omega_m$
    denoting the matching  frequency and the label X being either 
    EOB (for $a_5=40$, $v_{\rm pole}=0.5074$) or NR.}
 \end{center}
\end{figure}

Finally, we claim that, not only the {\it phase}, but also the {\it amplitude}
of the new, resummed EOB waveform Eq.~(\ref{eq:EOBh22}) exhibits a remarkable agreement
with the NR data of~\cite{arXiv:0710.0158}. Again, as Ref.~\cite{arXiv:0710.0158}
gave their results in the form of differences T4-NR, we plot in
Fig.~\ref{fig:fig7} the analog of the right panel of Fig.~19 there. We choose
again the ``optimum''  values $a_5=40$, $v_{\rm pole}=0.5074$ used in
Fig.~\ref{fig:fig6} and plot the NR~$\to$~EOB analogs of the curves plotted by
them in Fig.~19. Namely, we plot, at once, the $\omega_2$-, $\omega_3$- and
$\omega_4$-matched amplitude differences 
$\left[\Delta^{\omega_n}A/A\right]_{\rm T4EOB}=\left(A_{\rm
  T4}^{\omega_m}-A_{\rm EOB}\right)/A_{\rm EOB}$, where, as above, the T4 time is
shifted so that $\omega_{\rm T4}(t')$ and $\omega_{\rm EOB}(t)$ agree when
$\omega_{\rm EOB}(t_m)=\omega_m$. In addition, we plot, as empty circles, some
points taken (by approximate measurements of ours) from the corresponding curve 
$[\Delta^{\omega_4}A/A]_{\rm T4NR}$ plotted on the right panel of Fig.~19 
of~\cite{arXiv:0710.0158}. 
The remarkable visual agreement  between these empty circles and our 
$\left(\Delta^{\omega_4}A/A\right)_{\rm T4EOB}$ curve shows that: (i) the new,
resummed 3~PN amplitude introduced in Ref.~\cite{arXiv:0705.2519} and
defined in Eqs.~(\ref{eq:EOBh22})~(\ref{eq:delta22}) above agrees remarkably 
well with the NR one on the full time interval, 
$900M\leq t_{\rm EOB}\leq 3460M$, (ii) by contrast the
Taylor~T4~3.5/2.5 PN amplitude shows a significant disagreement ($\sim -8\%$)
in the same interval. Note that, though Ref.~\cite{arXiv:0710.0158} emphasizes
that the {\it non-resummed} 3~PN-accurate waveform of~\cite{arXiv:0710.0614} 
``improves agreement significantly'' compared to the 2.5~PN one (used above), this
improvement only concerns the early part of the inspiral. Indeed, Fig.~21 
of~\cite{arXiv:0710.0158} shows that the amplitude of  Taylor~T4~3.5/3.0 tends
again to diverge together with Taylor~T4~3.5/2.5 at the end of the inspiral:
i.e., we think, precisely around the ``dip'' (near $\omega_4$) exhibited in 
Fig.~\ref{fig:fig7} above.

\section{Conclusions}
\label{sec:conclusions}

We have investigated the agreement (in phase and in amplitude) between the
predictions of the Effective-One-Body (EOB) formalism and some accurate
numerical data. We used as numerical data both (as a warm up) some old results
on the energy flux from circular orbits of a test mass around a non spinning
black hole~\cite{gr-qc/9505030}, and some very recent results 
of the Caltech-Cornell group about the $\ell=m=2$ gravitational wave emitted 
by 15 orbits of an inspiralling system of two equal-mass non-spinning black 
holes~\cite{arXiv:0710.0158}.

In our warm up, test-mass example we showed how a slight tuning of the 
{\it flexibility parameter}~\cite{gr-qc/0211041} $v_{\rm pole}$ (away from the
naively expected value $v_{\rm pole}^{\rm standard}(\nu=0)=1/\sqrt{3}=0.57735)$
to the value $v_{\rm pole}^{\rm best}(\nu=0)\simeq 0.540$ allowed one to fit
remarkably well the flux function $F(v;\nu=0)$ during the full inspiral,
$0\leq v\leq v_{\rm LSO}=1/\sqrt{6}$.

In the comparable mass case ($\nu=m_1 m_2/(m_1+m_2)^2\sim 1/4$) we 
followed~\cite{arXiv:0705.2519} in introducing a new, resummed 
3~PN-accurate\footnote{Actually, our waveform has a greater accuracy than 3~PN 
                      in that it incorporates the test-mass limit of the 4~PN
		      and 5~PN amplitude corrections. We shall occasionally
		      refer to this PN accuracy as being $3^{+2}$-PN.}
EOB-type $\ell=m=2$ waveform. We then showed how to compute, for any values of
the EOB flexibility parameters $a_5$ (parametrizing 4~PN and higher
conservative orbital interactions) and $v_{\rm pole}$ (parametrizing $\nu$-dependent
4~PN and higher effects in the resummed radiation reaction) the EOB
predictions for the $\ell=m=2$ gravitational curvature wave
$\Psi^{22}_4\propto \partial^2_t h_{22}^{\rm EOB}\propto A_{\rm
  EOB}(t)e^{-{\rm i}\phi_{\rm EOB}(t)}$.

We then compared the EOB predictions for the gravitational wave (GW) phase,
$\phi_{\rm EOB}(t)$, and amplitude, $A_{\rm EOB}(t)$, to the numerical
relativity results of~\cite{arXiv:0710.0158}, say $\phi_{\rm NR}(t)$, $A_{\rm NR}(t)$,
using often as intermediary (as Ref.~\cite{arXiv:0710.0158}) the so-called
Taylor~T4~3.5/2.5 post-Newtonian predictions  $\phi_{\rm T4}(t)$, $A_{\rm T4}(t)$.
Our main conclusions are:

(i) In the GW frequency domain $M\omega<0.08$ where the Taylor~T4~3.5/2.5 phase
matches well with the NR phase, the EOB phase matches at least as well with
the NR phase. A good EOB/NR match is obtained both for the standard
``non-tuned'' EOB flexibility parameters 
$a_5=0$, $v_{\rm pole}=v_{\rm  pole}^{\rm DIS}(\nu)$ corresponding to our 
current {\it analytical} knowledge~\cite{Damour:2000we,gr-qc/9708034} and for
``tuned'' EOB flexibility parameters. 

(ii) For higher GW frequencies, $0.08<M\omega\lesssim 0.14$, while 
Taylor~T4~3.5/2.5 starts to significantly diverge from the NR phase, 
we showed that the standard ``non-tuned'' EOB phasing continues to
stay in phase with NR within $\pm8\times 10^{-3}$ GW cycles (see Fig.~\ref{fig:fig5}).
Moreover, one can calibrate $a_5$ and $v_{\rm pole}$ so that the 
EOB phase matches with the NR phasing to the truly remarkable level 
of $\pm 10^{-3}$ GW cycles over 30 GW cycles!

(iii) We proposed several ways of ``best fitting'' the $(a_5, v_{\rm pole})$-dependent
EOB predictions to accurate NR data: (a) by using the {\it intrinsic} representation
of the phase evolution given by the reduced phase-acceleration function
$a_{\omega}(\omega)$, Eq.~(\ref{eq:aomega}); (b) by using selected ratios
$\Delta^{\omega_m}\phi_{\rm T4EOB}/\Delta^{\omega_m}\phi_{\rm T4NR}$ and
constraining them to be close to unity; and (c) by using an $L_{\infty}$ 
norm of the difference between ($\omega_m$-matched) $\phi_{\rm EOB}(t)$ 
and $\phi_{\rm NR}^{\omega_m}\left(t'_{\omega_m}\right)$.

Our results are given in several Figures. Notably, Fig.~\ref{fig:fig3}
gives, for each given value of $a_5$, what is the optimum value of 
$v_{\rm  pole}$ which best fits (in the sense of the ratio 
$\rho_{\omega_4}^{\rm  bwd}$, Eq.~(\ref{eq:ratio4})) the NR data. 
Then, Fig.~\ref{fig:fig4} plots the $L_{\infty}$ distance  (on a large
time-interval roughly corresponding to the full simulation of~\cite{arXiv:0710.0158})
betweem $\phi_{\rm EOB}(t)$ and 
$\phi_{\rm  NR}^{\omega_4}\left(t'_{\omega_4}\right)$ as a function of $a_5$
(for $v_{\rm pole}=v_{\rm pole}(a_5)$ given by Fig.~\ref{fig:fig3}). We find
that the absolute value of the maximum dephasing between EOB and NR can 
be as small as 0.006 radians (or 0.001 GW cycles) 
if $30\lesssim a_5\lesssim 52$. 
However, it is difficult to be precise about the
``preferred'' valued of $a_5$. We recall in this respect that, recently,
Ref.~\cite{arXiv:0706.3732} has tried to constrain the value of $a_5$ 
(keeping, however, $v_{\rm pole}$ fixed to $v^{\rm DIS}_{\rm pole}(\nu)$, 
and without using our improved EOB waveform) by maximizing the overlap 
between EOB and NR plunge waveforms. 
They found that the overlap was good (and flat) over a rather large interval 
of values of $a_5$ (that they denote as $\lambda$), roughly centered around 
$a_5\simeq 60$. We note, however, that this behavior might be due 
(at least in part) to the phenomenon pointed out in~\cite{gr-qc/0211041}. 
In the latter reference (where $a_5$ was denoted as $b_5$), it was found that 
the use of EOB templates based on $a_5=50$ (rather than $a_5=0$) allowed one to 
have large overlaps (large ``effectualnesses'') with all other EOB templates. 
At this stage, we therefore do not have yet any precise knowledge of what 
might be the preferred ``effective'' value of $a_5$. 
Our work, however, shows that there is a quite strict correlation between the
best-fit choices of $a_5$ and $v_{\rm pole}$. When, in the future, $a_5$
becomes precisely known, it will be interesting to see what is the
corresponding value of $v_{\rm pole}(\nu=1/4)$ and to compare it to the
best-fit value $v_{\rm pole}(\nu=0)\simeq 0.540$ obtained in our warm-up Sec.~\ref{sec:poisson}.

For instance, the couple $a_5=40$, $v_{\rm pole}=0.5074$ yields a remarkable
good fit to the NR data reported in~\cite{arXiv:0710.0158}. We show the 
comparison of the various phasings (NR, EOB, T4) in Fig.~\ref{fig:fig6}. This 
Figure clearly exhibits how our best-fit EOB phase does a much better job than 
any non-resummed PN approximant at following the NR phase. We finally get 
dephasings smaller than $\pm 0.006$ radians (i.e. $< 10^{-3}$ GW cycles!) 
over about 30 GW cycles!

Finally, we exhibited in Fig.~\ref{fig:fig7} how the amplitude of our new,
resummed $3^{+2}$-PN-accurate EOB waveform, Eq.~(\ref{eq:EOBh22}), exhibits a 
remarkable agreement with the corresponding amplitude of the NR data 
of~\cite{arXiv:0710.0158}. The agreement is clearly better than any, 
non resummed PN amplitude, including the recent 3~PN-accurate one of 
Kidder~\cite{arXiv:0710.0614}.

We think that the present work, taken in conjunction with other recent works
on the EOB-NR comparison~\cite{arXiv:0706.3732}~\cite{arXiv:0704.3550,arXiv:0705.2519}, 
confirms the remarkable ability of the EOB formalism (especially in its
recently improved avatars) to agree with NR results. 
Note in particular that the level of phase agreement reached here is better by
a factor $30$ ($\pm0.001$ GW cycles versus $\pm0.03$ GW cycles for $\nu=1/4$) than what
was recently achieved, for merger signals, in Ref.~\cite{arXiv:0706.3732}
using less accurate versions of EOB waveforms than the 
one used  here. We suggest that the
ground-based interferometric GW detectors should include in their template
banks the new, extended and improved EOB waveforms which are being developed
and notably the resummed one introduced in~\cite{arXiv:0705.2519} and generalized
here. We also suggest that NR data be made available in some repository, 
soon after the first published results, to expert theorists willing to 
extract the physical information they contain.

\acknowledgments
We thank Eric Poisson for providing us with the numerical data of
Fig.~\ref{fig:fig1}, Larry Kidder for informative e-mail exchange 
about his 3~PN results and Bala Iyer for help in comparing our $\nu$-dependent
waveform to Kidder's result. The commercial software Mathematica$^{\rm TM}$ 
and Matlab$^{\rm TM}$ have been broadly used in the preparation of this paper.

\end{document}